\documentclass[preprint]{aastex}
\usepackage{graphicx}
\begin{document}

\singlespace

\title{The Unusual Object IC 2144/MWC 778}
\author{G. H. Herbig}
\affil{Institute for Astronomy, University of Hawaii, Honolulu, HI 96822;\\
herbig@ifa.hawaii.edu}
\and
\author{William D. Vacca}
\affil{SOFIA-USRA, NASA Ames Research Center, Moffett Field, CA 94035;\\
wvacca@sofia.usra.edu}

\begin{abstract}

IC 2144 is a small reflection nebula located in the zone of
avoidance near the Galactic anticenter.  It has been investigated here
largely on the basis of Keck/HIRES optical spectroscopy (R $\approx$ 48,000)
and a SpeX spectrogram of the near-IR (R = 2000) obtained at the NASA IRTF.  The
only star in the nebula that is obvious in the optical or near-IR is
 the peculiar emission-line object MWC 778
($V$ = 12.8), which resembles a T Tauri star in some respects.  What
appear to be  F- or G-type absorption features are detectable in its
optical region under the very
 complex emission line spectrum; their radial velocity agrees with the
CO velocity of the larger cloud in which IC 2144 is embedded.  There
are significant differences between the spectrum of the brightest area of
the nebula and of MWC 778, the presumed illuminator, an issue discussed
in some detail. The distance of IC 2144 is inferred to be about 1.0 kpc
by reference to other star-forming regions in the vicinity.  The extinction
is large, as demonstrated by $[$\ion{Fe}{2}$]$ emission line ratios in
the near-IR and by the strength of the diffuse interstellar band spectrum;
a provisional value of $A_{V}$ of 3.0 mag was assumed.  The SED of MWC 778
rises steeply beyond about 1 $\mu$m, with a slope characteristic of a
Class I source.  Integration of the flux distribution leads to an IR
luminosity of about 510 $L_{\sun}$.  If MWC 778 is indeed a F- or G-type
pre--main-sequence star several magnitudes above the ZAMS, a population of
faint emission H$\alpha$ stars would be expected in the vicinity.  Such
a search, like other investigations that are recommended in this paper,
has yet to be carried out.

\end{abstract}


\section{Introduction}
	On conventional photographs such as the Palomar Sky Atlas, IC 2144
appears as a roughly rectangular object about 25\arcsec $\times$ 16\arcsec
\ that might be mistaken for a small irregular galaxy like M82, or an \ion{H}{2}
region.  It is in fact a complex reflection nebulosity, apparently
illuminated by an off-center star of $V = 12.8$, and embedded in a small, rather
inconspicuous dark cloud.  That star\footnote{$\alpha$ = 5$^{h}$\, 
50$^{m}$\, 13$^{s}$.5,
$\delta$ = +23$^{\circ}$\, 52$'$\, 17$''$ (J2000); $l$ = 184$^{\circ}$.9,
$b$ = $-$1$^{\circ}$.7 }
was found by \citet{mer49} to have H$\alpha$ as well as lines of \ion{Fe}{2} and
[\ion{Fe}{2}] in emission.  They named it MWC 778 and assigned type Bpe.  In what
follows, we continue to distinguish between the star MWC 778 and the
nebulosity IC 2144. 

	IC 2144 is reproduced in Figure 1 from an exposure obtained with
the SuprimeCam instrument behind an H$\alpha$ filter at the Subaru Telescope\footnote{The 
Subaru Telescope is operated by the National Astronomical
Observatory of Japan.} and kindly made available by Bo
Reipurth. MWC 778 is located near the white spot in the overexposed nebulosity.
The nebula appears to be much larger (about 120\arcsec $\times$ 100\arcsec)
and more complex than apparent on earlier images. Figure 2 is the same
image plotted on a logarithmic intensity scale.  The internal structure is
shown in Figures 3a, b, and c, which are $R$-band images of IC 2144, extracted from
2 s snapshots (obtained in poor seeing) of the slit plane of the Keck I
HIRES spectrograph during one of the spectroscopic exposures to be described
later.  Very bright nebulosity immediately surrounds the star
and extends about 5\arcsec\ to the northwest (the image of MWC 778 itself
is slightly elongated in the same direction), where it splits into two curving
forks, best seen in Figure 3b, which is the same image as Figure 3a, on a
logarithmic intensity scale.  At longer wavelengths the surface brightness of
IC 2144 becomes so high and so structured as to confuse the 2MASS $J$, $H$, and
$K_{s}$ images, which have a resolution of about 4\arcsec.

	Optically, the appearance of the interior of IC 2144 is defined by
what appears to be a band of dust in the shape of a reversed question mark
that begins north of MWC 778, curves as it crosses the bright background of
the nebula to the east, and appears to unwind as it disappears against the
faintly luminous background south of MWC 778. There is a faint (R magnitude
 about 17) star (arrowed in Fig. 3b) about 4\arcsec\ northwest of MWC 778, at the
 beginning of this feature, and another about 3  \arcsec southwest.
It is unknown whether they are related to the nebula,  or if the brighter
of the two northeast forks is really the illuminated base of the dust band.

	\citet{allen74} noted the emission lines of H, \ion{Fe}{2},
 [\ion{Fe}{2}] and [\ion{S}{2}] in MWC 778, and performed the first
near-IR photometry out to 18 $\mu$m.  He commented that ``its almost linearly
rising energy distribution [in the $\lambda$F$_{\lambda}$, $\lambda$
 plane] is unique."  The IRAS fluxes (it is the point source 05471+2351
 and the Small Scale Structure Source X0547+238) reinforce that impression:
they have not begun to decline as far as the 100 $\mu$m point, although the
IRAS catalog notes that both the 60 and 100 $\mu$m fluxes are uncertain
because of the bright background.  Published photometry
of the star is collected in Table 1, but it is not always clear what aperture
size was used, or if allowance was made for a contribution by the
bright nebulosity.

	IC 2144 is near the Galactic anticenter, so the LSR velocity of its
associated CO (about +2 km s$^{-1}$: $\S$ 2) is too small to establish
a kinematic
distance.  On the sky, there is no well-defined obscuring cloud nearby.
Within a radius of 3$\arcdeg$--4$\arcdeg$ of IC 2144 there are two
 distant \ion{H}{2} regions
(S 242, at about 2 kpc, and S 243), two distant star clusters (NGC 2129 at 2
kpc and Be 21 at 5--6 kpc), and the bright B star 139 Tau (distant 0.9--1.1
kpc).  There is no reason to believe that IC 2144 is associated with any of
these objects.  There are two star-forming regions in that general
direction, but both are about 3$\arcdeg$ distant.  To the northwest
is the small cloud containing the HAeBe star RR Tau and a few T Tauri Stars (TTS); its distance
has been estimated as 380 pc by \citet{rost99}. The contrast
between the star density projected upon that cloud and that of its background
is large,  unlike the situation at IC 2144 where the contrast is small.
This is to be expected if IC 2144 is considerably more distant.  To the
south is the small cloud CB34, studied most recently by \citet{khan02}.  Its
distance is usually taken to be 1.5 kpc on the assumption that it is
associated with the Gem OB1 association, which lies still farther east.
On the basis of all these samples, we assume for the purposes of
this paper that the distance of IC 2144 is 1.0 kpc, but emphasize the
uncertainty of this estimate.

\section {Optical Spectroscopy}

	Spectrograms of the object were obtained on 2003 December 13 and 2004
November 21 with the HIRES spectrograph at the Keck I telescope.\footnote{The W. M. 
Keck Observatory is operated as a scientific
partnership among the California Institute of Technology, the University of
California, and NASA.  The Observatory was made possible by the generous
financial support of the W. M. Keck Foundation.  The spectrograph is
described in some detail in http://www2.keck/hawaii.edu/inst/hires/.}
The first series covers the region 4350--6740 \AA\ at a resolution of
 about 48,000
with slit dimensions of 0\farcs86 $\times$ 7$''$.  The seeing then was poor:
star images have a FWHM of about 1\farcs5.  The first exposure of that
series was
centered on MWC 778.
 The second crossed the brightest portion of the nebulous
arc about 5$''$ northwest of the star; its precise location is shown in Figure
3c.  The 2004 exposure was also centered on MWC 778, but covered 4400--8800
\AA\ at slightly higher resolution.  Reductions were carried out with standard
IRAF procedures.  The average S/N per pixel in the continuum of the two
(reduced) exposures of the star averaged about 50 in the 5000 \AA\ region,
 100 near 6200 \AA, and 130 near 7700 \AA.

	The spectrum of MWC 778 is dominated by a strong ($W \simeq 170$ \AA\ )
H$\alpha$ emission, followed by many emission lines of \ion{Fe}{2},
[\ion{Fe}{2}], [\ion{O}{1}], [\ion{S}{2}],  and \ion{Si}{2}, and a few 
of [\ion{Ni}{2}] and [\ion{Cr}{2}].  The profiles of H$\alpha$ and H$\beta$ are shown
in the upper section of Figure 4, with velocities\footnote{The radial
velocities in
this paper are heliocentric, unless noted otherwise.  The correction of
heliocentric to LSR at the position of IC 2144 is $-$11.8 km s$^{-1}$. } of
several features marked.  \citet{viei03}, at a resolution of 9,000, also noted
the off-center reversal in H$\alpha$ and the presence of [\ion{S}{2}] and
[O\, I]; they gave the type as ``B1?."  The spectrum between 3900 and
6700 \AA\ at moderate resolution was reproduced by \citet{su06}.  Only
emission lines are apparent. Their classification was ``BQ[\, ]." 

  	The lines of $[$\ion{S}{2}$]$ and $[$\ion{N}{2}$]$ $\lambda$6583 scatter in
velocity between +14 and +17 km s$^{-1}$, in agreement with $[$\ion{Fe}{2}$]$.
However, both [\ion{O}{1}] $\lambda$6300 and $\lambda$6363 are double, with
peak velocities of +13 and +29 km s$^{-1}$.  That region (on the 2004
spectrum) is shown in Figure 5, where the splitting of the $[$\ion{O}{1}] lines
is apparent (see the expanded profile of $\lambda$6300 in Fig. 5), as is
 the striking difference between the widths of
permitted and forbidden lines.  Despite their duplicity, the FWHM of the
combined $[$\ion{O}{1}] lines is only about 60 km s$^{-1}$, as compared to 220
km s$^{-1}$ for \ion{Si}{2} $\lambda$6347.
	  
	 The \ion{Fe}{2} lines are double, with a splitting of about 72 km s$^{-1}$.
This is most obvious in the weaker \ion{Fe}{2} lines such as $\lambda\lambda$
5234 and 5362, shown in Figure 6 (plotted on a velocity scale), while in the
strongest, such as $\lambda\lambda$ 4923 and 5018, it is apparent only as
unresolved structure in the shortward wing.  The [\ion{Fe}{2}] lines, on the other
hand, are narrow and single (see two examples in Fig. 6), at a mean velocity of
+16.5 $\pm$ 1 km s$^{-1}$, in clear contrast to \ion{Fe}{2} at $-28 \pm 2$ and
+44 $\pm$ 2 km s$^{-1}$. 

	The spectrum of MWC 778 is
cluttered with line emission, but an absorption-line spectrum is dimly
discernible in the clearer regions.  No real classification is possible,
but most of the detectable lines are those expected of an F- or G-type star,
although a late A type cannot be ruled out.
Those lines are shallow, probably because of veiling, and are not sharp;
their widths correspond to a $v$ sin $i$ of
30--40 km s$^{-1}$.  This is apparent in Figure 7,  a plot of the 6080--6160
\AA\ region in MWC 778 (top), compared to that region in the G0\, V star
HR 8314 (bottom), and (middle) the same HR 8314 spectrum broadened to
$v$ sin $i$ = 35 km s$^{-1}$.  Note that the vertical scales
in the several panels of Figure 7 are very different in order to display
each spectrum at approximately the same line amplitude.

	The Li\, I 6707 \AA\ line is present in MWC 778, as expected for a
pre--main-sequence star, at an equivalent width of 47 m\AA.  The velocity
of the absorption  spectrum, from 23 measurable lines, is +13 $\pm$ 2
km s$^{-1}$  The B-type absorption spectrum suspected by \citet{viei03} is
not in evidence.

       The CO cloud velocity at IC 2144 is +14.1 $\pm$ 0.3 km s$^{-1}$
according to \citet{blitz82} and +13.8 $\pm$ 0.1 according to \citet{wout89}.
In the optical region of MWC 778,  strong interstellar \ion{Na}{1} D$_{1,2}$
lines are present on the HIRES spectrograms (equivalent
widths 0.85 and 0.74 \AA ) at +13 km s$^{-1}$.  The one unblended interstellar
\ion{K}{1} line $\lambda$7698 is at +12 km s$^{-1}$.  Two sets of much weaker
IS lines fall in the shortward wings of the D lines, at  $-$4 and  $-$15
 km s$^{-1}$.
The diffuse interstellar band spectrum is strong; several diffuse interstellar bands (DIBs) are
indicated in Figure 5. Equivalent widths and velocities of some of the
 more prominent DIBs are given in Table 2; their mean velocity is
+12 $\pm$ 1 km s$^{-1}$. 

	The agreement of cloud CO and MWC 778 absorption line velocities
is as expected if the star formed within that cloud.  The agreement with
the optical interstellar velocities (\ion{Na}{1}, \ion{K}{1}) is compatible
with the idea that that material also is associated with the CO cloud,
although at that longitude only a weak dependence of velocity upon distance
is expected.

	The emission spectrum of MWC 778 is very similar to that of another
high-luminosity pre--main-sequence object, LkH$\alpha$-101, in which the
\ion{Fe}{2} lines also are double, the $[$\ion{Fe}{2}$]$ lines are single at
an intermediate velocity, the $[$\ion{O}{1}] lines are double, and the
\ion{Si}{2} lines are very broad \citep{herb04}.  In that case, the
\ion{Fe}{2} splitting was interpreted as produced in a tipped annulus
in rotation around a 15 $M_{\sun}$ star.

\section{ Near-Infrared Spectroscopy}

	MWC 778 was observed at the NASA Infrared Telescope Facility on
Mauna Kea on 2005 October 21, 17:30 UT with SpeX, the facility near-infrared
medium-resolution cross-dispersed spectrograph \citep{ray03}. Eight individual
exposures, each lasting 90 s, were obtained using the short-wavelength
cross-dispersed (SXD) mode of SpeX. This mode yields spectra spanning
the wavelength range 0.8--2.5 $\mu$m divided into six spectral orders. The
observations were acquired in ``pair mode," in which the object was observed
at two separate positions along the 15\arcsec-long slit. The slit width
was set to 0$\farcs$3, which yields a nominal resolving power of 2000 for
the SXD spectra. The slit was set to the parallactic angle during the
observations. The airmass was about 1.08. Observations of an A0$\, $V star,
used as a ``telluric standard" to correct for absorption due to Earth's
atmosphere and to flux calibrate the target spectra, were obtained
immediately preceding the observations of MWC 778. The airmass difference
between the observations of the object and the standard was 0.05. A set of
internal flat fields and arc frames was obtained immediately after the
observations of MWC 778 for flat fielding and wavelength calibration
purposes.

	The data were reduced using Spextool \citep{cus04}, the IDL-based
package developed for the reduction of SpeX data. The Spextool package
performs nonlinearity corrections, flat fielding, image pair subtraction,
aperture definition, optimal extraction, and wavelength calibration. The
eight sets of spectra resulting from the individual exposures were median
combined and then corrected for telluric absorption and flux calibrated
using the extracted A0\,V telluric standard spectra
and the technique and software described by \citet{vac03}. The
spectra from the individual orders were then spliced
together by matching the flux levels in the overlapping wavelength
regions, and regions of poor atmospheric transmission
were removed. The final spectrum is shown in Figure 8, and in expanded
sections in Figure 9. The
signal-to-noise ratio varies across the spectral range,
but is of the order of 100 at the shortest wavelength and rises to
several hundred at the center of the $K$ band.  Table 3 is a list of the
wavelengths, identifications, fluxes, and equivalent widths of emission
 lines measured on this spectrogram.

	The errors of the line fluxes and equivalent widths were estimated
following the procedure of \citet{cus05}, which uses the errors of the
individual pixel values as produced by Spextool \citep{cus04,vac04}, and
follows the method of \citet{sem92}.

	The photometric magnitudes of MWC 778 (source 4
in Table 1) were estimated by
multiplying the SpeX spectrum by the MKO filter transmission functions
\citep{tok02} and integrating over the passbands.

\section{Electron Density and Temperature}

	The multitude of emission lines present in the NIR spectrum of
MWC 778 allow, in principle, the physical parameters of the line-emitting
regions to be determined via the comparison of observed line
strength ratios with theoretical values. Of course, the
observed line strengths are affected by extinction, and to properly
take this into account, an iterative procedure is required.
Nevertheless, there are several relevant line ratios available on
which extinction has a relatively minor effect, and that 
can be used to estimate the electron density ($n_{e}$) and
temperature ($T_{e}$).

	The [\ion{Fe}{2}] lines in the $J$ and $H$ bands provide three line
ratios that can be used for this purpose.  \citet{bau95} used the \ion{Fe}{2}
model of \citet{rz93} to compute the ([\ion{Fe}{2}] 1.298 + 1.295 $\mu$m)/[\ion{Fe}{2}] 1.257 
and [\ion{Fe}{2}] 1.533/[\ion{Fe}{2}] 1.257 $\mu$m ratios as a
function of $n_{e}$.  A combination of these results and the observed
([\ion{Fe}{2}] 1.298 + 1.295)/[\ion{Fe}{2}] 1.257 $\mu$m line ratio of 0.2 produces a
 nearly reddening-independent estimate of $\log n_{e}$ [cm$^{-3}$]
 $\sim$ 4.15. To obtain the
second line ratio from the observed spectrum, the contribution of Br 18
to the observed Br 18 + [\ion{Fe}{2}] 1.533 blend was estimated by interpolating
the Br line series strengths as a function of quantum number on either
side of the line. After subtraction of the
Br 18 contribution, it was found that [\ion{Fe}{2}] 1.533/[\ion{Fe}{2}] 1.257 $\sim$
0.2, which leads to a value of $\log n_{e}$ [cm$^{-3}$] $\sim$
4.3.  However, this ratio is affected by extinction, which drives the
ratio and the corresponding derived electron density to larger
values, so this value is an upper limit. The results of \citet{rz93} also
provide a third estimate of $n_{e}$; by combination of their theoretical
[\ion{Fe}{2}] 1.533/[\ion{Fe}{2}] 1.644 ratio with our (Br 18-subtracted) value,
an electron density of $\log n_{e}$ [cm$^{-3}$] $\sim$ 3.8 is obtained.
Again, this ratio is affected by extinction, but in the opposite sense of
that for the [\ion{Fe}{2}] 1.533/[\ion{Fe}{2}] 1.257 ratio. Hence the derived density
value is a lower limit. In light of the uncertainties in the
observed line ratios, all three density estimates are consistent with a
value of $\log n_{e}$ [cm$^{-3}$] $\sim$ 4.1.

	A limit on $T_{e}$ in the [\ion{Fe}{2}] emitting region can be placed by
using the theoretical predictions of \citet{rz93} for
the [\ion{Fe}{2}] 8617 \AA/[\ion{Fe}{2}] 1.257 $\mu$m ratio. For an assumed electron
density of $\log n_{e}$ [cm$^{-3}$] $\sim$ 4.1, the observed ratio
corresponds to $T_{e} \sim 8000$ K; a lower limit as reddening will reduce
the observed line ratio.

\section{The Spectrum of the Nebulosity}

	One expects that the spectrum of a reflection nebulosity
illuminated by MWC 778 would mimic that of the star.  That is not the case.
The slit of the second HIRES exposure crossed the brightest area of
IC 2144 west and northwest of the star, as shown in Figure 3c.  The spectrum of
the nebula was extracted at four positions along that 7$''$-long slit; the
boundaries of these segments are marked in that figure.  A sky sample was
taken from the end of the slit above segment D, and subtracted from each
segment. There is a considerable variation in both emission lines and
continuous background along the slit, the brightness of both being highest
at segment B.  However, the two do not maintain a fixed ratio as would be
expected if the same illuminating source were being scattered at all
positions along the slit. The effect is shown in Table 4, where the
brightness (pixel$^{-2}$) of [O\, I] $\lambda$6300, averaged over its
central 0.24 \AA, is compared to the brightness of the nearby continuum,
both expressed in units of the continuum brightness in segment B.
The [\ion{O}{1}] lines are double in both star and nebula, but at
different velocities: +13 and +29 km s$^{-1}$ in the star, +1 and +13
km s$^{-1}$ in the nebula; see Table 5 for details.

	 There is a striking difference between star and nebula 
in the profiles of H$\alpha$ and H$\beta$.  Figure 4
shows the two lines in the star (above) and in the nebula (below), the latter
being the sum of the spectra extracted from the four slit segments.  Both lines
in the nebula are structureless and nearly symmetric, while what appears
to be a shortward absorption reversal in the star is absent in the nebula. 
Furthermore, the line
peaks in the nebula are displaced 10--20 km s$^{-1}$ shortward with respect
to their positions in the star.  These differences, as well as a less obvious
mismatch in the longward wing, are apparent when the profile of H$\alpha$ in
the star is superposed upon the profiles of that line from the four individual
samples of the nebular spectrum: see Figure 10, and the details in Table 5.

	The lines of [\ion{Fe}{2}] that are so prominent in MWC 778 are
essentially absent (with respect to \ion{Fe}{2}) in the nebula.  The evidence is
in Table 6, where the equivalent widths of (mostly) unblended lines of \ion{Fe}{2}
 and [\ion{Fe}{2}{]} in star and in nebula are compared.

	Table 7 lists the $W$'s and radial velocities of the $[$\ion{S}{2}$]$
 6717, 6730 \AA\ lines in MWC 778 (on both dates) and at several points in
the nebula (in 2003), and electron densities estimated by comparing the
observed [\ion{S}{2}] 6717/6730 equivalent width ratios with intensity ratios
 computed with a 5-level atom model as a function of $n_e$ and $T_e$.
 (An electron temperature of $10^4$ K was assumed, but there is relatively
 little variation in the
line ratios with temperature.)  

\section{Interpretation of the Mismatch between Star and Nebula:}

\subsection{ A Travel-Time Effect? }

	One explanation of the differences between star and nebula might be
that they result from a travel-time effect.  That is, could the spectrum
of MWC 778 have changed before that change was replicated in the nebular
spectrum?  The projected separation between the star and the point in the
nebula where that spectrum was obtained is about 5$\arcsec$, which at 1 kpc
corresponds to a light travel time of 29/cos$\phi$ days, $\phi$ being the
inclination of the radius vector to the plane of the sky.  Variations in the
emission-line spectra of pre--main-sequence stars are of course well known,
but in the case of MWC 778 there is no evidence of change in the shortward
structure of H$\alpha$ between the two HIRES observations, 344 days apart.
More observations would be instructive, but at the moment this remains a
somewhat unlikely possibility.

\subsection{ Or Another Illuminator of the Nebula? }

	Another possibility is that the nebula might be illuminated to a
significant degree, not by MWC 778, but by some other source that is deeply
obscured as seen from our direction.  If so, at IR wavelengths the intensity
peak might move away from the optical position of MWC 778.  To explore this
 possibility,
IRAC images of the field (at 3.6, 4.5, 5.8 and 8.0 $\mu$m) were extracted from
the Spitzer archive.\footnote{ These observations were made with the Spitzer
Space Telescope, which is operated by the Jet Propulsion Laboratory,
California Institute of Technology, under a contract with NASA.} 
About ten stars whose coordinates are available
from the USNO A2 compilation lie within about 3$\arcmin$ of MWC 778, and
 at least six of these were measureable on all the IRAC images and served to
 define a common ($\alpha$, $\delta$) coordinate system in which the
intensity peak of each image could be located.  It was found that there was
 no systematic shift in its position between 0.65 and 8.0 $\mu$m; the total
scatter around
the mean position was 0$\fs$03,\ 0$\farcs$4.  The faint star arrowed in
Figure 3b does not appear in the Spitzer images, and thus is of no
concern in this connection.

	Polarization imagery, by discovering the direction from which the
nebula is illuminated, could settle the question of whether there is a source
other than MWC 778. The polarization maps of the nebula presented by
\citet{per06} do not suggest that there is another source.  One has to say
that if there were another illuminator of the nebula, it would be surprising
that its spectrum is so similar, almost in detail, to that of MWC 778.  A
useful additional test would be to determine if the F-G absorption lines of
MWC 778 are present also in the nebular spectrum.  Unfortunately,
the present spectrum of the nebula is too noisy (S/N $\approx$ 15 in
the continuum) for that check.

\subsection{ Or MWC 778 Seen from Different Directions?  }

	There remains the more interesting possibility, rather
than the nebula being illuminated by some other source, that MWC 778 is
in fact the illuminator but is not spherically symmetric, and so looks quite
different as seen from our direction and from any point in the dust that
 it illuminates.

	Although the profile of H$\alpha$ remains about the same through
the sequence of nebular samples A through D, its equivalent width decreases
with (projected) distance from the star.  The [\ion{O}{1}] lines are double
in both star and nebula, but at different velocities, velocity separations,
and equivalent widths; the details for both H$\alpha$ and $\lambda$6300
are given in Table 5.

        The $\lambda$6300 line is seen in many of the TTS and HAeBe stars
observed by \citet{acke05}.  Although a variety of line profiles was
found, $\lambda$6300 is double in some objects, notably in two of the three
HAeBe stars studied by \citet{vand08}, although in
those stars the peak intensity of the line was only 0.05 to 0.10 of the
continuum intensity, as compared with 2 to 2.5 in MWC 778.  Nevertheless,
their interpretation of the doubling as the consequence of emission in an
unresolved, inclined disk might apply to MWC 778.  We pursue that
possibility.

	Imagine that the rotational axis of the central star embedded
in the image of MWC 778 is inclined to the line of sight, is directed
to the northwest, and that it is surrounded by an equatorial disk in
Keplerian rotation.  Then the double \ion{Fe}{2} lines (splitting 72 km s$^{-1}$)
could be formed in the disk near the star, while the double [\ion{O}{1}]  lines
(splitting 16 km s$^{-1}$) would be formed farther out.  The low $n_{e}$'s
inferred from [\ion{S}{2}] and [\ion{Fe}{2}] and the fact that lines of both ions
are single indicate that they originate still farther out in the
disk where the orbital velocity is too small for their splitting to be
resolved.  The different sites thus inferred for [\ion{S}{2}] and [\ion{O}{1}] 
may be a consequence of the latter being favored nearer the star by a
special mechanism, namely the UV photodissociation of OH and H$_{2}$O
(see \citealt{acke05} for references).  Then the ``reversed question mark"
feature in the nebula, which resembles the one-armed dust spirals seen
at some pre--main-sequence stars like SU Aur, may be material unwound
from the disk periphery.  If the flux of the outflowing wind that
produces the shortward absorption reversals at the Balmer lines is
latitude dependent, then the strength of that feature in the nebular
spectrum would depend upon the location of the viewing point in the nebula.

        If the [\ion{O}{1}]  splitting is indeed due to disk rotation,
the unequal peak intensities (Table 5) would mean that the [\ion{O}{1}]  brightness
is nonuniform in disk azimuth.  Then the difference in doublet splitting
between star (16 km s$^{-1}$) and nebula (about 11 km s$^{-1}$), and the
change in doublet intensity ratio between star and nebula would be
explained if the nebula ``sees'' the disk at a different inclination and
from a different direction.  The shortward velocity shift of the doublet
midpoint and of the Balmer lines from star to nebula would then be
explained by motion of nebula with respect to star.

        Unexplained is the peculiar shortward structure
at H$\alpha$ and H$\beta$ in the star.  The emission wings of those lines
would have to be extended about 2 \AA\ shortward, with respect to the
nebular profiles, to provide a background if that is indeed an 
absorption reversal.  Another possibility is that it is not
an absorption at all, but only a gap between the main emission
peak and a secondary emission peak at  $-$127 km s$^{-1}$ at H$\alpha$ or
at  $-$145 km s$^{-1}$ at H$\beta$.  Yet no such feature at those
displacements is present in the spectrum of MWC 778 at the \ion{He}{1}, 
 \ion{Fe}{2}, or unblended Paschen lines (P14, P17, P20).

	Future observations should clarify these issues.

\section{Extinction}

    	The extinction as inferred from the optical colors will be discussed
later, but other indicators also are available.

        (1) Normally, the numerous Paschen and Brackett emission lines
in this spectrum would provide excellent diagnostics of the reddening
of MWC 778.  However, as is apparent in Figure 11, the relative line
strengths do not vary in a manner consistent with Case B.  Specifically,
ratios of the strengths of higher order Pa lines to Pa$\beta$, and Br
lines to Br$\gamma$ appear to be far too strong compared with the Case B
predictions of \citet{hum87} for all but the most extreme
combinations of electron temperature and density: $T_{e}$ $<$ 3000 K
and 10$^{7}$ $<$ n$_{e}$ $\leq$ 10$^{10}$ cm$^{-3}$.  Probably
the lower members of the Pa and Br series are optically
thick (or somehow otherwise suppressed relative to the higher
levels). Our measured Pa$\beta$/Br$\gamma$ line
ratio agrees quite well with the ``Case C" values given
by \citet{xu92}, which applies when both the Lyman and Balmer
line series are optically thick. In this case, one
would expect that some of the lowest lying Pa and Br lines should
also be optically thick (Storey, private communication).
Unfortunately, \citet{xu92} did not give theoretical ratios for
any other combination of lines seen in our spectra.

 	An attempt was made to fit the relative fluxes of
various sets of higher order H lines in both the Br and Pa
series, allowing the electron density, electron temperature, and
reddening to vary. No consistent value of the reddening
could be found that resulted in acceptable fits to all of the lines
considered. In fact, when the analysis was restricted 
to only the higher order ($n >$ 12) Br line fluxes, the best fits yielded
negative reddening values.

	Next, [\ion{Fe}{2}] transitions arising from the same upper level
were examined.  The intrinsic flux ratios of such pairs of lines should
follow from their wavelengths and $A$ values. The
strongest suitable line pair observed in MWC 778 consists of the
$a^6D_{9/2} - a^4D_{7/2}$ 1.257 $\mu$m and the $a^4F_{9/2} - a^4D_{7/2}$
1.644 $\mu$m transitions. (Other [\ion{Fe}{2}] lines are present but either lie
in a region of poor telluric correction, or are blended, or are very weak.)
Their observed line fluxes, the $A$ values given by \citet{nuss88}, and
the extinction law of \citet{rie85} result in a color excess of $E(B-V)
 = 0.75 \pm 0.06$, or $A_{V} = 2.31 \pm 0.19$.

	(2) As remarked, the interstellar diffuse band spectrum is very
strong; several DIBs are indicated in Figure 5.  The equivalent widths (and
velocities) of some prominent features are given in Table 2, together with
the average $W$'s of the same bands in five of the B-type stars in the
obscured cluster NGC 1579 \citep{herb04}.  Obviously, the
DIBs are substantially stronger in MWC 778 than in those B stars, by a
factor 1.3 on the average.  Since the average $A_{V}$ of those stars is
3.5 mag, the implication is that the extinction of MWC 778 is still larger. 
But it is well known that although DIB strengths increase with extinction
in a general sense, there is significant scatter about any regression line,
there are substantial local variations, and DIB strengths do not respond
to the heavy extinction of dense molecular clouds or of circumstellar material.

	The looseness of the correlation of DIB strength with color
excess is illustrated by the relationship between the equivalent width
of the $\lambda$5780 DIB and $E(B-V)$ from \citet{herb93}.  The
$W$($\lambda$5780) of MWC 778 from Table 2 leads to $E(B-V) = 0.83$, or
$A_{V}$ = 2.57, but the values for individual stars near that value of $W$
spread between $E(B-V)$ = 0.63 and 1.6, or $A_{V}$ = 1.95 and 5.0.

	Given the scatter of these indicators, we can say only
that for MWC 778, $A_{V}$ may be about 3 mag.

\section{Luminosity}

	Figure 12 shows the SED of IC 2144/MWC 778, constructed from the
data of Table 1.  Solid points represent the observed data points, while
open circles show the result of correction for $A_{V} = 3.0$, under the
assumption of normal reddening (to 10 $\mu$m from \citealt{rie85},
longward
of 10 $\mu$m from \citealt{mat90}).  The slope of the SED between 2.0 and 25
$\mu$m, $\alpha = d \log(\lambda F_{\lambda})/d \log \lambda$,
is +0.7.  If taken at face value this indicates Class I membership.

	The energy contained in the infrared excess between 2 and 100 $\mu$m
can be estimated by integrating the flux densities $F_{\lambda}$ between
these limits.  If it is assumed that $A_{V}$ = 3.0 mag and that the distance
is 1 kpc, the resulting IR luminosity is $370 \pm 40\ L_{\sun}$, where that
uncertainty reflects only the assumed 30\% uncertainties in the IRAS
60 $\mu$m and 100 $\mu$m flux densities. An estimate of the additional 
excess beyond 100 $\mu$m can be made by assuming that it follows a Planck
distribution peaking at that wavelength \citep{coh73, chav81}.
This adds about 140 $L_{\sun}$ to the above value. As pointed out by the referee, such luminosities assume that the object radiates isotropically.

	Note that the optical-region fluxes include not only the
near-point-source MWC 778, but at longer wavelengths a larger fraction
of IC 2144.

	Consider what can be learned about MWC 778 by the color of that
steeply rising section of the SED shortward of about 1 $\mu$m.   The solid
curve in Figure 12, passing through the open circles as far as 0.366 $\mu$m,
is the Planck function for 15,000 K, corresponding to a mid-B-type star.
That is quite incompatible with the absorption spectrum.  The discrepancy
is somewhat smaller if one considers instead the ($V-I_{C}$) color, thus
avoiding the contribution of the strong emission lines in the blue and
near-ultraviolet.
The contributions of line emission to the V and I$_{C}$ magnitudes,
obtained by summing the equivalent widths of all measureable
lines across the V and I$_{C}$ passbands, are estimated as 0.03 mag  and
0.05 mag, respectively, so the corrected $V$ and ($V-I_{C}$) are 12.83
and +1.48.  Table 8 gives the main-sequence spectral types \citep{bes88}
corresponding to this color if
normal reddening is assumed, and  correction is applied for several values
of $A_{V}$.  Absolute magnitudes $M_{V}$ follow if $V$ is
corrected for those $A_{V}$'s and distances of 0.7, 1.0, and 1.3 kpc.

	But this suggests a mid-A spectral type for the star,  still
in conflict with the actual classification of F or G (although as explained,
that type is very uncertain because of heavy interference by the overlying
emission spectrum).  Either of the following two assumptions could 
 reduce this disagreement.

(1) The foregoing hinges on the assumption of normal reddening, i.e., that
 $R_{V} = A_{V}/(B-V) = 3.1$.  Larger values of $R_{V}$ are often
found in dense clouds, so consider the possibility that here $R_{V}$ has the
not unreasonable value of 5.0.  The consequent value of the extinction
correction can be estimated from the relationships provided by
 \citet{card89} from which $A_{\lambda} / A_{V}$ can be obtained as a
function of $\lambda$ and $R_{V}$.  These, weighted by the filter
transmission and integrated across the  $V$ and $I_{C}$ passbands, result in 
$A_{V} / E(V-I_{C}$) values of 2.44 for $R{_V}$ = 3.1, and 2.96 for 5.0.
The inferred spectral type for $A_{V}$ = 3.0 mag then moves from
A5\,V (Table 8) to F3.5\,V, a better match for the actual classification.

(2) The discrepancy would be lessened also if $A_{V}$ is smaller than has been
assumed, if (for example) $A_{V}$ = 2.0 instead of 3.0 mag, then the region
shortward of 1 $\mu$m would conform to a 9000 K Planck function (early A type)
and the ($V-I_{C}$) color to type F9\ V (Table 8). 

	Neither assumption can be ruled out at this time.  But 
the correct explanation of this discrepancy between color and spectral type
may turn out to be very simple: that an object with a bizarre emission-line
spectrum like MWC 778 cannot be expected to have the colors and energy
distribution of a normal ZAMS star.

\section{Final Remarks}

	How is MWC 778 to be classified?  Figure 13 shows the location of about
70 of the pre-main sequence stars of the Tau-Aur clouds in the (B-V),(U-B)
plane, from the compilation of \citet{ken95}.  Points in boxes are classical T 
Tauri Stars (CTTS) having $W$(H$\alpha$) $>$ 60 \AA\ (CTTS), solid points 
are CTTS having $W$(H$\alpha)$ between
10 and 60 \AA, and small open circles are weak-line T Tauri stars (WTTS; $W$(H$\alpha)$ $<$
 10 \AA).  A large cross marks the location of MWC 778. Clearly, it falls among
the CTTS, as would be expected from its large W(H$\alpha$),  but it is
earlier in type than most of the TTS in that area of Figure 13, and it is
atypical also in the prominence of the $[$\ion{Fe}{2}$]$ spectrum and other
forbidden lines as $[$\ion{Ni}{2}$]$ $\lambda\lambda$ 7377, 7412 and $[$\ion{Cr}{2}$]$
$\lambda$8000, and the weakness of the IR \ion{Ca}{2} lines.  Somewhat similar
emission line patterns can be found among the stars observed by \citet{ham89}
and \citet{ham94}, which included LkH$\alpha$-101.  Certainly, MWC 778 is of
much higher luminosity than conventional TTS, as shown particularly by its
SED.

	If the MWC 778 point in Figure 13 is translated to the upper left
along the reddening vector by $A_{V} = 3.0$ mag, it approaches the ZAMS in 
the early B region, as would be expected from the fit to the dereddened
SED shortward of 1 $\mu$m described above.

	On the basis of the evidence presented here, MWC 778 is an F- or
G-type pre--main-sequence star several magnitudes above the ZAMS.  It is not
a T Tauri star, but might be considered a later-type analog of the HAeBe stars.

	Our picture (\S\ 6.3) envisages disk structure concealed within the
unresolved image of MWC 778.  \citet{per06} has suggested that there is
a much larger disk also illuminated by MWC 778 that is outlined by
the curved arc extending about 6$\arcsec$ north of the star, and by
its shorter extension to about 4$\arcsec$ south.  We have been referring
to this structure as ``the nebula."  Our discussion of its spectrum
is based on a sample taken at the base of the northern arc about
3$\arcsec$--5$\arcsec$ west and northwest of the star (Fig. 3c).
Perrin's idea could be tested by high-resolution CO spectroscopy of
the entire area.

	We stress that the investigation of IC 2144, its contents, and its
neighborhood has just begun, and that the values of distance and extinction
that we have adopted are provisional. A closer examination of the absorption
spectrum of MWC 778 is recommended.  Very desirable would be a search
for the faint H$\alpha$-emission stars that would be expected in and around
IC 2144 if MWC 778 is in fact the most massive member of such a population.
At longer wavelengths, high-resolution CO mapping of the nebula could be
very worthwhile, for example because it is not obvious whether the 'reversed
question mark' seen against the bright background of IC 2144 (Figs. 2 and 3)
represents a real feature or is simply the shadow of structure
near MWC 778 thrown on that dust screen.  An examination at longer IR
wavelengths and in the submillimeter of MWC 778 is also desirable.

\acknowledgements

	G. H. was partially supported during this investigation by the U.S.
National Science Foundation under grants AST 02-04021 and AST 07-02941.  We
are grateful for access to the SIMBAD database at CDS, Strasbourg, France, 
to Colin Aspin and Bo Reipurth for comments, and to Marshall Perrin for
making his thesis available and for a valuable comment.

\singlespace

\newpage

\begin{deluxetable}{lclcc|lccc}
\tabletypesize{\small}
\tablecolumns{8}
\tablewidth{0pt}
\tablecaption{Published Photometry of MWC 778}
\tablehead{
\colhead{Color} &\colhead{Source} & \colhead{Magnitude}
 &\colhead{F$_{\lambda}$\tablenotemark{a}} & & \colhead{Color} &\colhead{Source}
 & \colhead{Magnitude} & \colhead{F$_{\lambda}$\tablenotemark{a} } }

\startdata
 $U$ & 1  & 13.45  & \nodata & & $L$ & 3  & 6.3             & \nodata \\
 $B$ & 1  & 13.72  & \nodata & & [8.28]    & MSX     & \nodata         & \phn\phn3.86    \\
 $V$  & 1  & 12.80  & \nodata & & [8.6$] $    & 3  & 2.9  & \nodata \\
 $R_{c}$ & 1 & 11.98  & \nodata & & [11.3]    & 3  & 1.8  & \nodata \\
 $I_{c}$ & 1 & 11.30  & \nodata & & 12 $\mu$m       & IRAS    & \nodata         & \phn\phn6.40    \\
 $J$ & 2 & 10.24 $\pm$ 0.11  & \nodata & & [12.13]   & MSX     & \nodata         & \phn\phn5.36   \\
 $J$ & 4 & 10.14            & \nodata & & [14.65]   & MSX     & \nodata         & \phn\phn6.01   \\
 $H$ & 3  & \phn8.65            & \nodata & & [18]      & 3  & $-$0.3\phs & \nodata \\
 $H$ & 2 & \phn8.96 $\pm$ 0.03   & \nodata & & [21.3]    & MSX     & \nodata         & \phn36.61  \\
 $H$ & 4  & \phn9.15            & \nodata & & 25 $\mu$m       & IRAS    & \nodata         & \phn40.36  \\
 $K$ & 3  & \phn7.84            & \nodata & & 60 $\mu$m      & IRAS    & \nodata         & \phn\phm{:}97.49: \\
 $K$ & 2  & \phn8.02 $\pm$ 0.02  & \nodata & & 100 $\mu$m      & IRAS    & \nodata         & \phm{:}133.79: \\
 $K$ & 4  & \phn8.08  &  \nodata& &  &  & & \\ 
\enddata
\tablecomments {Sources: (1) \citealt{viei03};
(2) \citealt{gar97};
(3) \citealt{allen74};
(4) this paper.}
\tablenotetext{a}{The flux densities (in Janskys) are from the MSX (Midcourse Space Experiment) and IRAS
 Point Source Catalogs.  The IRAS values are not color corrected. }

\end{deluxetable}

\begin{deluxetable}{cccc}
\tabletypesize{\small}
\tablecolumns{4}
\tablewidth{0pt}
\tablecaption{DIBs in MWC 778 and NGC 1579 B Stars}
\tablehead{
\colhead{ }    & \multicolumn{2}{c}{MWC 778} & \colhead{B Stars}  \\
\cline{2-3} 
\colhead{Wavelength} & \colhead{$W$ } & \colhead{$v_{\sun}$ }
   & \colhead{$W$} \\
\colhead{(\AA ) } & \colhead{(m\AA)} & \colhead{(km s$^{-1}$)} &
 \colhead{(m\AA)} } 

\startdata
 5780.55 & 406.        &  +12. & \nodata \\
 5797.08 & 176.        &  \phn+8. & 101.  \\
 5849.81 &  \phn60.        &  +13. &  \phn42.  \\
 6195.99 &  \phn47.        &  +12. &  \phn41.  \\ 
 6379.22 &  \phn76.        &  +14. &  \phn58.  \\
 6613.63 & 190.        &  \phn+9. & 176.  \\
 6699.28 &  \phn22.        &  \phm{:}+11.:&  \phn19.  \\

\enddata

\end{deluxetable}  

\clearpage

\begin{deluxetable}{lllrlcll}
\tabletypesize{\small}
\tablecaption{Identifications, Fluxes and Equivalent Widths for Lines in
 the NIR Spectrum of MWC 778}
\tablecolumns{8}
\tablewidth{0pc}
\tablehead{
\colhead{$\lambda_{\rm lab}$\tablenotemark{a}} & \colhead{} &
\colhead{$\lambda_{\rm obs}$\tablenotemark{b}} &
\colhead{Flux} & \colhead{$\sigma$\tablenotemark{c}} & \colhead{$W_\lambda$}
 & \colhead{$\sigma$\tablenotemark{c}} & \colhead{} \\
\colhead{($\mu$m)}  & \colhead{Identification} & \colhead{($\mu$m)} & \multicolumn{2}{c}
{($10^{-17}~{\rm W ~m}^{-2}$)} & \colhead{(\AA)} & \colhead{(\AA)} &
 \colhead{Note} }
\startdata
0.822897 & \ion{Fe}{2} ($e^6D_{7/2} - 5p^4F^o_{9/2}$) & 0.82318  &  6.44  &  0.34  &  $-$2.30  &  0.12  & \nodata   \\
0.8288      & \ion{Fe}{2} ($e^6D - 5p^6F^o$)                     & 0.82898  &  5.25  &  0.28  &  $-$1.88  &  0.10  & \nodata   \\ 
                  & ?                                                                & 0.83052  &  3.78  &  0.32  &  $-$1.35  &  0.11  &  broad  \\
0.834555 & \ion{H}{1} (Pa23)                                                & 0.83475  &  1.35  &  0.18  &  $-$0.50  &  0.07  & \nodata   \\
0.835723 & \ion{H}{1} (Pa22) + \ion{Fe}{2} ($e^6D_{5/2} - 5p^6F^o_{5/2}$)     & 0.83601  &  3.60  &  0.23  &  $-$1.34  &  0.09  & \nodata   \\
0.837447 & \ion{H}{1} (Pa21)                                                & 0.83763  &  2.89  &  0.24  &  $-$1.09  &  0.09  & \nodata   \\
0.839240 & \ion{H}{1} (Pa20)                                                & 0.83947  &  4.08  &  0.25  &  $-$1.61  &  0.10  & \nodata   \\
0.841332 & \ion{H}{1} (Pa19)                                                & 0.84172  &  7.19  &  0.38  &  $-$2.84  &  0.15  & \nodata   \\
0.84465   & \ion{O}{1} ($3s^3S^o_1 -  3p^3P$) + \ion{H}{1} (Pa18)   &  0.84474 &  60.65 &  0.45  &  $-$24.73\phn  &  0.21& \nodata   \\
0.846726 & \ion{H}{1} (Pa17)                                                & 0.84695  &  11.44  &  0.36  &  $-$4.77  &  0.15 & \nodata   \\
0.850248  & \ion{H}{1} (Pa16) + \ion{Fe}{2} ($e^6D_{3/2} - 5p^6F^o_{5/2}$) & 0.85014 &  20.00  &  0.46  &  $-$8.55  &  0.20 &  blend  \\
                   & + Ca II ($3d^2D_{3/2} - 4p^2P^o_{3/2}$)?   &                 &               &            &            &               & \\
0.854539 & \ion{H}{1} (Pa15) + Ca II ($3d^2D_{5/2} - 4p^2P^o_{3/2}$)? & 0.85469  &  15.49  &  0.34  &  $-$6.77  &  0.15  & \nodata   \\
0.859839 & \ion{H}{1} (Pa14)                                                 & 0.86006  &  15.24  &  0.31  &  $-$6.73  &  0.14  & \nodata   \\
0.861696 & [\ion{Fe}{2}] ($a^4F_{9/2} - a^4P_{5/2}$)      &  0.86179  &   2.24  &  0.20  &  $-$0.99  &  0.09  & \nodata   \\
0.866502 & \ion{H}{1} (Pa13) + Ca II ($3d^2D_{3/2} - 4p^2P^o_{1/2}$)?  & 0.86668  &  18.20  &  0.34  &  $-$7.99  &  0.15  & \nodata   \\
0.868028 & \ion{N}{1} ($3s^4P_{5/2} - 3p^4D^o_{7/2}$) ?  & 0.86838  &    3.44  &  0.25  &  $-$1.51  &  0.11  &   wing \\
0.875047 & \ion{H}{1} (Pa12)                                                 & 0.87530  &  18.63  &  0.35  &  $-$8.16  &  0.15  & \nodata   \\
0.886279 & \ion{H}{1} (Pa11)                                                 & 0.88654  &  22.33  &  0.38  &  $-$9.73  &  0.17  & \nodata   \\
0.892665 & \ion{Fe}{2} ($e^4D_{7/2} - 5p^4D^o_{5/2}$) & 0.89289  &    4.52  &  0.24  &  $-$1.96  &  0.10  & \nodata   \\
0.901491 & \ion{H}{1} (Pa10)                                                 & 0.90172  &  22.87  &  0.37  &  $-$9.90  &  0.16  & \nodata   \\
0.907549 & \ion{Fe}{2} ($e^4D_{5/2} - 5p^4P^o_{5/2}$) &  0.90777  &   4.65  &  0.25  &  $-$2.00  &  0.11  & \nodata   \\
0.909483 & \ion{C}{1} ($3s^3P^o_2 - 3p^3P_2$) & 0.90978  &   3.56  &  0.26  &  $-$1.53  &  0.11  & \nodata   \\
                   & + \ion{Fe}{2} ($e^6D_{9/2} - 5p^6D^o_{7/2}$) ?               &                   &             &            &              &            & \\
0.912292 & \ion{Fe}{2} ($e^4D_{7/2} - 5p^4D^o_{7/2}$) & 0.91253 &  4.11  &  0.24  &  $-$1.76  &  0.10  &   blend \\
0.913238 & \ion{Fe}{2} ($e^4D_{7/2} - 5p^4F^o_{9/2}$)  & 0.91343  &  3.81  &  0.23  &  $-$1.63  &  0.10  &   blend \\
0.917805 & \ion{Fe}{2} ($e^4D_{3/2} - 5p^4F^o_{5/2}$)  & 0.91797  &  7.36  &  0.31  &  $-$3.14  &  0.13  & \nodata   \\
                   & + \ion{Fe}{2} ($e^4D_{5/2} - 5p^4F^o_{7/2}$)  &              &            &            &              &           &    \\
0.920313 & \ion{Fe}{2} ($e^4D_{1/2} - 5p^4F^o_{3/2}$)  & 0.92051  &  7.53  &  0.25  &  $-$3.21  &  0.11  &  wing  \\
0.922901 & \ion{H}{1} (Pa9) + \ion{Mg}{2} ($4s^2S_{1/2} - 4p^2P^o_{3/2}$) & 0.92276  &  38.56  &  0.38  &  $-$16.44\phn  &  0.17  &  blend/  \\
                  & +  [\ion{Fe}{2}] ($a^4F_{5/2} - a^4P_{3/2}$)? &            &           &             &             &           &  broad\\
0.924427 & \ion{Mg}{2} ($4s^2S_{1/2} - 4p^2P^o_{1/2}$) & 0.92473  &  8.74  &  0.21  &  $-$3.75  &  0.09  &  wing  \\
0.926599 & \ion{O}{1} ($^5P_3 - ^5D^o_4$)                       & 0.92680  &  2.99  &  0.17  &  $-$1.28  &  0.07  & \nodata   \\
0.9297      & \ion{Fe}{2} ($e^4D_{5/2} - 5p^4D^o_{5/2}$) & 0.93011  &  3.04  &  0.20  &  $-$1.30  &  0.09  & \nodata   \\
                   & + \ion{Fe}{2} ($e^6D_{9/2} - 5p^6D^o_{9/2}$) ? &           &            &            &             &            &     \\
0.93889   & \ion{Fe}{2} ($z^6D^o_{5/2} - c^4P_{3/2}$) ? & 0.93920  &  5.73  &  0.22  &  $-$2.46  &  0.09  &   blend  \\
                  & \ion{N}{1} ($3s^2P_{1/2} - 3p^2D^o_{3/2}$) ? &                &            &            &             &            &     \\
0.940573 & \ion{C}{1} ($3s^1P^o_1 - 3p^1D_2$)            & 0.94078  &  2.64  &  0.18  &  $-$1.13  &  0.08  & \nodata   \\
                   & + \ion{Fe}{2} ($e^4D_{7/2} - 5p^6F^o_{9/2}$) ? &          &           &            &              &            & \\
0.942843 & \ion{Fe}{2} ($e^4D_{7/2} - 4p^4G_{9/2}$)   & 0.94304  &  1.08  &  0.15  &  $-$0.46  &  0.07  & \nodata   \\
0.949779 & \ion{Fe}{2} ($z^4P^o_{5/2} - d^4P_{3/2}$)  & 0.94999   &  2.27  &  0.17  &  $-$0.97  &  0.07  & \nodata   \\
0.954598 & \ion{H}{1} (Pa~$\epsilon$)                              & 0.95485  &  32.72  &  0.25  &  $-$14.04\phn  &  0.11  & \nodata   \\
0.957258 & \ion{Fe}{2} ($z^4P^o_{5/2} - d^4P_{5/2}$)  & 0.95743  &  6.37  &  0.17  &  $-$2.73  &  0.07  & \nodata   \\
0.961494 & \ion{Fe}{2} ($z^6D^o_{1/2} - c^4P_{3/2}$)  &  0.96183  &   2.31  &  0.17  &  $-$0.99  &  0.07  & \nodata   \\
0.981210 & \ion{Fe}{2} ($z^4P^o_{3/2} - d^4P_{3/2}$)  &  0.98147  &   2.59  &  0.14  &  $-$1.09  &  0.06  & \nodata   \\
0.985026 & [\ion{C}{1}] ($^3P_2 - ^1D_2$)                       &  0.98521 &  1.76  &  0.15  &  $-$0.73  &  0.06  & \nodata   \\
                  & ?                                                               &  0.98824 & 1.26  &  0.11  &  $-$0.52  &  0.05  & \nodata   \\
0.989481 & \ion{Fe}{2} ($z^4P^o_{3/2} - d^4P_{3/2}$)  &  0.98969 &  2.31  &  0.13  &  $-$0.95  &  0.05  & \nodata   \\
0.991028 & \ion{Fe}{2} ($z^6D^o_{3/2} - c^4P_{1/2}$)  & 0.99128  &  1.78  &  0.12  &  $-$0.73  &  0.05  & \nodata   \\
0.995628 & \ion{Fe}{2} ($z^4F^o_{9/2} - b^4G_{9/2}$)  & 0.99584  &  3.04  &  0.13  &  $-$1.25  &  0.06  & \nodata   \\
0.999756 & \ion{Fe}{2} ($z^4F^o_{9/2} - b^4G_{11/2}$) & 1.00001  &  29.78  &  0.17  &  $-$12.19\phn  &  0.07  & \nodata   \\
1.004937 & \ion{H}{1} (Pa~$\delta$)                                  & 1.00521  &  43.45  &  0.21  &  $-$17.83\phn  &  0.09  & \nodata   \\
1.011248 & \ion{N}{1} ($3p^4D^o_{5/2} - 3d^4F_{7/2}$) ? & 1.01136  &  1.53  &  0.15  &  $-$0.63  &  0.06  & \nodata   \\
                  &   ?                                                             & 1.01327  &  1.07  &  0.14  &  $-$0.44  &  0.06  & \nodata   \\
1.017349 & \ion{Fe}{2} ($z^4D^o_{7/2} - b^4G_{9/2}$) & 1.01763  &  8.43  &  0.20  &  $-$3.49  &  0.08  & \nodata   \\
1.028673 & [S II] ($^2D^o_{3/2} - ^2P^o_{3/2}$) & 1.02891  &  1.35  &  0.14  &  $-$0.56  &  0.06  & \nodata   \\
1.032049 & [S II] ($^2D^o_{5/2} - ^2P^o_{3/2}$) & 1.03232  &  1.71  &  0.14  &  $-$0.71  &  0.06  & \nodata   \\
1.033641 & [S II] ($^2D^o_{3/2} - ^2P^o_{1/2}$) & 1.03383  &  0.70  &  0.12  &  $-$0.29  &  0.05  & \nodata   \\
1.0398      & [\ion{N}{1}] ($^2D^o_{5/2} - ^2P^o$)             & 1.04009 &  1.37  &  0.19  &  $-$0.57  &  0.08  &   blend \\
1.0407      & [\ion{N}{1}] ($^2D^o_{3/2} - ^2P^o$)             & 1.04089  &   1.40  &  0.18  &  $-$0.58  &  0.07  &  blend  \\
1.043254 & [\ion{Fe}{2}] ($a^2H_{9/2} - b^2G_{9/2}$) ?  & 1.04375   &   3.37  &  0.15  &  $-$1.40  &  0.06  & \nodata   \\
1.045969 & [Ni II]  ($^2F_{7/2} - ^4P_{5/2}$) ?       & 1.04613  &    1.33  &  0.14  &  $-$0.55  &  0.06  & \nodata   \\
1.049930 & \ion{Fe}{2}   ($z^4F^o_{7/2} - b^4G_{5/2}$)  & 1.05024  &  26.60  &  0.21  &  $-$11.00\phn  &  0.09  & \nodata   \\
1.052512 & \ion{Fe}{2} ($z^4D^o_{5/2} - b^4G_{7/2}$)   & 1.05279  &    4.55  &  0.16  &  $-$1.88  &  0.07  & \nodata   \\
1.054648 & \ion{Fe}{2} ($e^6D_{9/2} - y^6F^o_{11/2}$)  & 1.05484  &    2.15  &  0.14  &  $-$0.89  &  0.06  & \nodata   \\
1.0686     & \ion{C}{1} ($3s^3P^o - 3p^3D$)                        & 1.06898  &    5.01  &  0.17  &  $-$2.04  &  0.07  &  broad  \\
1.071106 & \ion{Fe}{2} ($e^6D_{3/2} - 4p^6D^o_{5/2}$)  & 1.07131  &    1.76  &  0.15  &  $-$0.72  &  0.06  & \nodata   \\
1.0830      & He I ($2s^3S - 2p^3P^o$)                     & 1.08350  &  19.67  &  0.29  &  $-$8.06  &  0.12  & \nodata   \\
1.086265 & \ion{Fe}{2} ($z^4F^o_{5/2} - b^4G_{7/2}$)    & 1.08660  &  23.47  &  0.26  &  $-$9.63  &  0.11  & \nodata   \\
1.09130   & He I ($3d^3D - 6f^3F^o$) ?                   & 1.09165  &    6.07  &  0.21  &  $-$2.49  &  0.08  &  wing \\
                  & + \ion{Mg}{2} ($3d^2D - 4p^2P^o_{3/2}$)? &                 &              &             &             &           &         \\
1.093810 & \ion{H}{1} (Pa~$\gamma$) + \ion{Mg}{2} ($3d^2D_{3/2} - 4p^2P^o_{1/2}$)?  & 1.09416  &  71.26  &  0.31  &  $-$29.31\phn  &  0.13  & \nodata   \\
1.112557 & \ion{Fe}{2} ($z^4F^o_{3/2} - b^4G_{5/2}$)     & 1.11285  &  16.59  &  0.25  &  $-$6.79  &  0.10  & \nodata   \\
1.1287      & \ion{O}{1} ($3p^3P - 3d^3D^o$)                       & 1.12898  &  31.88  &  0.28  &  $-$13.03\phn  &  0.11  & \nodata   \\
1.1754      & \ion{C}{1} ($3p^3D - 3d^3F^o$)                       & 1.17554  &   3.27  &  0.15  &  $-$1.33  &  0.06  & \nodata   \\
                   & ?                                                                 & 1.23289  &  0.59  &  0.10  &  $-$0.24  &  0.04  & \nodata   \\
1.23836    & \ion{Fe}{2} ($z^6F^o_{9/2} - c^4F_{7/2}$)     & 1.23877  &  0.67  &  0.11  &  $-$0.27  &  0.04  & \nodata   \\
1.24303    & \ion{Fe}{2} ($z^6F^o_{9/2} - c^4F_{9/2}$)     & 1.24343  &  0.86  &  0.10  &  $-$0.34  &  0.04  & \nodata   \\
                   & ?                                                             & 1.24703  &  2.71  &  0.14  &  $-$1.08  &  0.06  & broad   \\
1.256680 & [\ion{Fe}{2}] ($a^6D_{9/2} - a^4D_{7/2}$)  & 1.25713  &  9.57  &  0.13  &  $-$3.79  &  0.05  & \nodata   \\
                  & ?                                                              & 1.26009  &  1.74  &  0.11  &  $-$0.69  &  0.04  & \nodata   \\
1.278776 & [\ion{Fe}{2}] ($a^6D_{3/2} - a^4D_{3/2}$)  & 1.27916  &  2.06  &  0.10  &  $-$0.81  &  0.04  &  wing  \\
1.281808 & \ion{H}{1} (Pa~$\beta$)                                   & 1.28208  &  118.51  &  0.20  &  $-$46.29\phn  &  0.08  & \nodata   \\
1.294268 & [\ion{Fe}{2}] ($a^6D_{5/2} - a^4D_{5/2}$)  & 1.29465  &  1.45  &  0.11  &  $-$0.56  &  0.04  & \nodata   \\
1.3164      & \ion{O}{1} ($3p^3P - 4s^3S^o_1$)               & 1.31671  &  22.88  &  0.15  &  $-$9.04  &  0.06  & \nodata   \\
1.320554 & [\ion{Fe}{2}] ($a^6D_{7/2} - a^4D_{7/2}$)  & 1.32089  &  2.25  &  0.11  &  $-$0.89  &  0.04  & \nodata   \\
1.327776 & [\ion{Fe}{2}] ($a^6D_{3/2} - a^4D_{5/2}$)  & 1.32812  &  0.95  &  0.12  &  $-$0.38  &  0.05  & \nodata   \\
                   &   ?                                                           & 1.34361  &  0.83  &  0.12  &  $-$0.33  &  0.05  & \nodata   \\
1.493773 & \ion{H}{1} (Br26)                                               & 1.49400  &  0.86  &  0.22  &  $-$0.34  &  0.09  & \nodata   \\
1.496734 & \ion{H}{1} (Br25)                                               & 1.49720  &  1.05  &  0.21  &  $-$0.42  &  0.08  & \nodata   \\
1.500086 & \ion{H}{1} (Br24)                                               & 1.50051  &  1.56  &  0.24  &  $-$0.62  &  0.10  & \nodata   \\
1.503904 & \ion{H}{1} (Br 23)                                              & 1.50420  &  1.14  &  0.21  &  $-$0.46  &  0.08  &  abs \\
1.508278 & \ion{H}{1} (Br22)                                               & 1.50864  &  2.64  &  0.28  &  $-$1.06  &  0.11  & \nodata   \\
1.513322 & \ion{H}{1} (Br21)                                               & 1.51368  &  3.15  &  0.26  &  $-$1.27  &  0.10  & \nodata   \\
1.519184 & \ion{H}{1} (Br20)                                               & 1.51944  &  3.80  &  0.19  &  $-$1.53  &  0.08  & \nodata   \\
1.526054 & \ion{H}{1} (Br19)                                               & 1.52626  &  4.14  &  0.18  &  $-$1.68  &  0.07  & \nodata   \\
1.534179 & \ion{H}{1} (Br18) + [\ion{Fe}{2}] ($a^4F_{9/2} - a^4D_{5/2}$)  &   1.53428  &  6.74  &  0.22  &  $-$2.74  &  0.09  & \nodata   \\
1.543892 & \ion{H}{1} (Br17)                                               & 1.54426  &  5.84  &  0.20  &  $-$2.37  &  0.08  & \nodata   \\
1.555645 & \ion{H}{1} (Br16)                                               & 1.55602  &  6.42  &  0.21  &  $-$2.60  &  0.09  & \nodata   \\
1.570066 & \ion{H}{1} (Br15)                                               & 1.57052  &  7.77  &  0.23  &  $-$3.16  &  0.09  & \nodata   \\
                  &   ?                                                            & 1.57580  &  3.41  &  0.17  &  $-$1.39  &  0.07  & \nodata   \\
1.588054 & \ion{H}{1} (Br14)                                               & 1.58817  &  8.30  &  0.18  &  $-$3.36  &  0.08  & \nodata   \\
1.599473 & [\ion{Fe}{2}] ($a^4F_{7/2} - a^4D_{3/2}$)   & 1.59995  &  1.64  &  0.14  &  $-$0.66  &  0.06  & \nodata   \\
1.610931 & \ion{H}{1} (Br13)                                               & 1.61132  &  10.60  &  0.19  &  $-$4.29  &  0.08  & \nodata   \\
1.640719 & \ion{H}{1} (Br12)                                               & 1.64103  &  10.06  &  0.17  &  $-$4.04  &  0.07  & \nodata   \\
1.643550 & [\ion{Fe}{2}] ($a^4F_{9/2} - a^4D_{7/2}$)   & 1.64397  &   8.98  &  0.14  &  $-$3.60  &  0.06  & \nodata   \\
1.680652 & \ion{H}{1} (Br11) +  [\ion{Fe}{2}] ($a^4F_{7/2} - a^4D_{5/2}$)  & 1.67992  &  20.66  &  0.26  &  $-$8.28  &  0.10  &  blend/  \\
                   & + \ion{Fe}{2} ($z^4F^o_{9/2} - c^4F_{7/2}$)  &               &             &            &             &            & broad\\
1.687320  & \ion{Fe}{2} ($z^4F^o_{9/2} - c^4F_{9/2}$) & 1.68771  &  17.35  &  0.20  &  $-$6.95  &  0.08  & \nodata   \\
                  & ?                                                              & 1.69859  &  1.10  &  0.16  &  $-$0.44  &  0.07  & \nodata   \\
1.736211 & \ion{H}{1} (Br10) + \ion{Fe}{2} ($z^4D^o_{7/2} - c^4P_{5/2}$)     & 1.73618  &  13.94  &  0.22  &  $-$5.61  &  0.09  &  blend  \\
1.741401 & \ion{Fe}{2} ($z^4D^o_{7/2} - c^4F_{7/2}$) & 1.74176  &  5.64  &  0.18  &  $-$2.27  &  0.07  & \nodata   \\
1.744934 & [\ion{Fe}{2}] ($a^4F_{3/2} - a^4D_{1/2}$) ? & 1.74562  &  1.04  &  0.13  &  $-$0.42  &  0.05  & \nodata   \\
1.755168 & \ion{Fe}{2} ($z^4D^o_{7/2} - c^4F_{5/2}$) ? & 1.75598  &  0.87  &  0.15  &  $-$0.35  &  0.06  & \nodata   \\
1.944556 & \ion{H}{1} (Br~$\delta$)                                   & 1.94487  &  15.61  &  0.46  &  $-$6.38  &  0.19  & \nodata   \\
1.957204 & \ion{Fe}{2} ($z^4F^o_{5/2} - c^4F_{7/2}$)  & 1.95761  &  4.99  &  0.27  &  $-$2.05  &  0.11  & \nodata   \\
1.974611 & \ion{Fe}{2} ($z^4F^o_{5/2} - c^4F_{5/2}$) ? & 1.97513  &  5.13  &  0.28  &  $-$2.13  &  0.12  & \nodata   \\
                  & ?                                                              & 1.98775  &  1.06  &  0.24  &  $-$0.44  &  0.10  & \nodata   \\
                  & ?                                                              & 2.00180  &  2.21  &  0.19  &  $-$0.91  &  0.08  & \nodata   \\
2.015123 & [\ion{Fe}{2}] ($a^2G_{9/2} - a^2H_{9/2}$) ?  & 2.01571  &  0.77  &  0.18  &  $-$0.32  &  0.07  & \nodata   \\
2.046005 & [\ion{Fe}{2}] ($a^4P_{5/2} - a^2P_{3/2}$)   & 2.04647  &  0.56  &  0.10  &  $-$0.24  &  0.04  & \nodata   \\
2.060025 & \ion{Fe}{2}  ($z^4F^o_{5/2} - c^4F_{5/2}$)  & 2.06023  &  2.72  &  0.12  &  $-$1.15  &  0.05  & \nodata   \\
                   & + He I ($2s^1S_0 - 2p^1P^o_1$) ? &                   &             &            &             &            &   \\
2.088810 & \ion{Fe}{2}  ($z^4F^o_{3/2} - c^4F_{3/2}$) ?   & 2.08942  &  4.52  &  0.13  &  $-$1.92  &  0.05  & \nodata   \\
                  & ?                                                              & 2.11798  &  0.27  &  0.09  &  $-$0.11  &  0.04  & \nodata   \\
2.121254 & H$_2$ (1,0)S(1)                                   & 2.12182  &  0.81  &  0.08  &  $-$0.35  &  0.04  & \nodata   \\
2.165529 & \ion{H}{1} (Br~$\gamma$)                              & 2.16590  &  24.44  &  0.19  &  $-$10.57\phn &  0.08  & \nodata   \\
                  & ?                                                              & 2.22401  &  0.68  &  0.13  &  $-$0.30  &  0.06  & \nodata   \\
                  & ?                                                              & 2.24107  &  0.52  &  0.10  &  $-$0.23  &  0.05  & \nodata   \\

\enddata

\tablecomments{ ``blend" = obvious blend of multiple components;
``broad" = conspicuously broader than other lines;
``wing" = line is in the wing of a much stronger neighbor;
``abs" = line appears to be sitting in an absorption trough;
``?" = identification uncertain. }

\tablenotetext{a}{\,Laboratory wavelength in air of strongest component.}
\tablenotetext{b}{Observed vacuum wavelength.}
\tablenotetext{c}{Statistical uncertainties. Absolute errors in both flux and equivalent width are expected to be $~15$\%.}

\end{deluxetable}
\clearpage

\begin{deluxetable}{cccc}
\tabletypesize{\small}
\tablecolumns{4}
\tablewidth{0pt}
\tablecaption{Relative Brightnesses of [O\, I] $\lambda$6300 and Continuum
in the Nebula}
\tablehead{
\colhead{Segment} & \colhead{$\lambda$6300} &\colhead{Continuum}
 &\colhead{Ratio} }

\startdata
 A & 2.71  & 0.64  & 4.23  \\
 B & 4.58  & 1.00  & 4.58  \\
 C & 3.01  & 0.85  & 3.54  \\
 D & 1.41  & 0.53  & 2.66  \\
\enddata

\end{deluxetable}  

\begin{deluxetable}{lll|llc|lc|cll}
\tabletypesize{\small}
\tablecolumns{10}
\tablewidth{0pt}
\tablecaption{H$\alpha$ and [\ion{O}{1}] $\lambda$6300 in MWC 778 and
 the IC 2144 Nebulosity }
\tablehead{
 \colhead{} & \multicolumn{2}{c}{H$\alpha$} &  & \multicolumn{6}{c}{$\lambda$6300} \\
\cline{2-3} \cline{5-10}  
 \colhead{} & \colhead{$W$} & \colhead{$v_{\sun}$ } & &
 \colhead{$W$} & \colhead{$v_{\sun}$} & 
 \colhead{$W$} & \colhead{$v_{\sun}$} &   
 \colhead{$W$} & \colhead{$v_{\sun}$} \\
\colhead{Source} & \colhead{(\AA)} & \colhead{(km s$^{-1}$)} & & 
\colhead{(\AA)} & \colhead{(km s$^{-1}$)} &
\colhead{(\AA)} & \colhead{(km s$^{-1}$)} &   
\colhead{(\AA)} & \colhead{(km s$^{-1}$)}  }

\startdata
Star 2003 Dec   & 164. & + 73.\tablenotemark{a} & &
 \nodata & \nodata &\  0.72  & +13. 
 & 0.70  & +30.  \\ 
Star 2004 Nov   & 154. & + 68.\tablenotemark{b} & &
 \nodata & \nodata &\  0.74  & +13.          
 & 0.61      & +29. \\ 
Nebula at A      & 136. & + 57.  & &
 0.35  &  + 1.&\  0.66       & +13.         
 & \nodata   & \nodata \\  
Nebula at B      & 137. & + 50. & &
  0.27      &  + 3.&\  0.97       & +13. 
 & \nodata   & \nodata \\ 
Nebula at C      & 116  & + 52. & &
  0.36      &  + 2.&\  0.90       & +13. 
 & \nodata   & \nodata \\ 
Nebula at D      & 103. & + 54.: & &
  0.4:      &  + 4.&\ 0.4:       & +12. 
 & \nodata   & \nodata \\ 
Nebula 3$\arcsec$ NW of star  & 154.  & + 49. & &
 \nodata &\nodata &\  1.27  & +14. 
 & \nodata   & \nodata \\

\enddata

\tablenotetext{a}{ The main peak of the H$\alpha$ emission line is at
 +73. km s$^{-1}$, the shortward minor peak is at  $-$115. km s$^{-1}$, and
 the interpeak minimum is at  $-$64. km s$^{-1}$. }
\tablenotetext{b}{ The main peak of the H$\alpha$ emission line is at
 + 68. km s$^{-1}$, the shortward minor peak is at  $-$98. km s$^{-1}$, and
the interpeak minimum is at  $-$59. km s$^{-1}$. }
\end{deluxetable}  

\clearpage
\begin{deluxetable}{lllcc|lllcc}
\tabletypesize{\small}
\tablecolumns{8}
\tablewidth{0pt}
\tablecaption{Comparison of \ion{Fe}{2} and $[$\ion{Fe}{2}$]$ Lines in MWC 778 and
IC 2144}
\tablehead{
\colhead{ } & \colhead{Wavelength } &\colhead{ } &\colhead{$W$ (star)} &\colhead{$W$ (nebula)} &   
\colhead{ } & \colhead{Wavelength } &\colhead{ } &\colhead{$W$ (star)} &\colhead{$W$ (nebula)}\\
\colhead{Ion} & \colhead{(\AA)} &\colhead{RMT} &\colhead{(\AA)} &\colhead{(\AA)} &                   
\colhead{Ion} & \colhead{(\AA)} &\colhead{RMT} &\colhead{(\AA)} &\colhead{(\AA)} }

\startdata                    

$[$\ion {Ni}{2}$]$ & 6666.82 & \nodata & 0.21  & $<$0.03\phm{$<$}  & $[$\ion{Fe}{2}$]$ & 5158.78 & 19F & 1.57 &  0.21 \\
 \ion {Fe}{2} & 6456.39  &     74  & 1.3\phn  &  1.12 & $[$\ion{Fe}{2}$]$ & 5111.63      & 19F &    0.14 &  $<$0.02 \\
 \ion {Fe}{2} & 6416.92  &    74   & 0.22 &  \phd0.25:& \ion{Fe}{2}       & 5018.43      &  42 &     4.1 &   3.55 \\
 \ion{Fe}{2} & 6317.99  & \nodata & 1.62 &  0.88 & $[$\ion{Fe}{2}$]$ & 4973.39      & 20F &    0.22 & $<$0.02  \\
 \ion{Fe}{2} & 6247.54  & 74 & 0.72 &  0.54 & $[$\ion{Fe}{2}$]$ & 4950.74      & 20F &    0.20 & $<$0.02  \\
 \ion{Fe}{2} & 5835.49  & \nodata & 0.10 &  \phd0.09:& $[$\ion{Fe}{2}$]$ & 4947.37      & 20F &    \phd0.13:& $<$0.02  \\
$[$\ion {Fe}{2}$]$ & 5746.97 &  34F& 0.20 & $<$0.02\phm{$<$} & \ion{Fe}{2}       & 4923.93      & 42  &    3.1  &  3.4 \\
 \ion{Fe}{2} & 5534.83  & \nodata & 0.37 &  0.25 & $[$\ion{Fe}{2}$]$ & 4905.34      & 20F &    0.38 & $<$0.1 \\
$[$\ion{Fe}{2}$]$ & 5527.34 &  17F& 0.31 & $<$0.03\phm{$<$} & $[$\ion{Fe}{2}$]$   & 4874.48      & 20F &    0.18 &  \phd0.04: \\
$[$\ion{Fe}{2}$]$ & 5433.13 &  18F& \phd0.29:& $<$0.03\phm{$<$} & $[$\ion{Fe}{2}$]$ & 4814.53      & 20F &    0.97 &  0.16 \\
$[$\ion{Fe}{2}$]$ & 5376.45 &  19F& 0.54 &  \phd0.04:& $[$\ion{Fe}{2}$]$ & 4774.72      & 20F &    0.19 & $<$0.04 \\
 \ion{Fe}{2} & 5362.87  &  49     & 0.46 &  0.24 & $[$\ion{Fe}{2}$]$ & 4639.67      &  4F &    0.15 & $<$0.02 \\
 \ion{Fe}{2} & 5316.62  &  49     & \phd2.7:\phn &  \phd2.6: & \ion{Fe}{2}       & 4629.34      &  37 &     1.00&  0.55  \\
$[$\ion{Fe}{2}$]$ & 5296.83 &  19F& 0.09 & $<$0.03\phm{$<$} & \ion{Fe}{2}       & 4583.83      &  38 &     1.62&  0.86  \\
$[$\ion{Fe}{2}$]$ & 5261.62 &  19F& 0.75 & $<$0.02\phm{$<$} & \ion{Fe}{2}       & 4555.89      &  37 &     0.9:&  1.3  \\
 \ion{Fe}{2} & 5234.62  &  49     & 1.15 &  1.0  & \ion{Fe}{2}       & 4549.47      &  38 &     1.15&  1.6 \\
$[$\ion{Fe}{2}$]$ & 5220.06 &  19F& 0.16 & $<$0.05\phm{$<$} & \ion{Fe}{2}       & 4515.34      &  37 &     0.52&  0.52 \\
$[$\ion{Fe}{2}$]$ & 5181.95 &  18F& 0.12 & $<$0.02\phm{$<$} &            &              &     &

\enddata
\end{deluxetable}

\begin{deluxetable}{lllclll}
\tabletypesize{\small}
\tablecolumns{7}
\tablewidth{0pt}
\tablecaption{$[$\ion{S}{2}$]$ Lines in MWC 778 and the IC 2144 Nebulosity }
\tablehead{
\colhead{ } & \multicolumn{2}{c}{$\lambda$6717} &\colhead{ } & \multicolumn{2}{c}{$\lambda$6730} & \colhead{ } \\
\cline{2-3} \cline{5-6}
\colhead{} & \colhead{$W$} & \colhead{$v_{\sun}$ } &\colhead{ } &
 \colhead{$W$} & \colhead {$v_{\sun}$ } & \colhead{}\\
 \colhead{Source} & \colhead{(m\AA)} & \colhead{(km s$^{-1}$) } &\colhead{ } &
 \colhead{(m\AA)} & \colhead {(km s$^{-1}$) } & \colhead{log $n_{e}$\tablenotemark{a}} }

\startdata
Star 2003 Dec   & \ 52.  & +16.  & & 118.:\tablenotemark{b}
  & +13.  & \ \ \  4.8:  \\
Star 2004 Nov  & \ 47.       & +13. & & 106.:\tablenotemark{b} & +16. &
\ \ \  4.7: \\
Nebula at A      & 128.        & +14.:& & 145.        & +10.  &  \ \ \ 3.0 \\
Nebula at B      & 185.        & +11. & & 240.        & +11.  &  \ \ \ 3.2 \\
Nebula at C      & 330.        & \ \  0. & & 388.        & + 1.  &  \ \ \ 3.0 \\
Nebula 3$\arcsec$ NW of star & 145.     & +11. & & 188. & + 9.  &  \ \ \ 3.2 \\ 
\enddata

\tablenotetext{a}{  $T_{e}$ = 10$^{4}$ K is assumed.}
\tablenotetext{b}{ In the star, the emission line at 6730 \AA\ is a blend of
[\ion{S}{2}] 6730.87 and [\ion{Fe}{2}] 6729.856. The [\ion{Fe}{2}] contribution was
removed by assuming it had the same profile as [\ion{Fe}{2}] 6809.226, of the same
multiplet, as measured on the 2004 exposure.}

\end{deluxetable}

\begin{deluxetable}{ccccc}
\tabletypesize{\small}
\tablecolumns{5}
\tablewidth{0pt}
\tablecaption{Inferred Spectral Type and $M_{V}$ of MWC 778 }
\tablehead{
\colhead{ } & \colhead{Sp. type} & \colhead{ } & \colhead{$M_{V}$} & \colhead{ } \\
\colhead{} & \colhead{having} & \colhead{  } & \colhead{if} & \colhead{ } \\
\colhead{$A_{V}$} & \colhead{$V-I_{C}$ } & \colhead{  } & \colhead{distance (kpc) is} &\colhead{ } \\
\cline{3-5}
\colhead{(mag) } & \colhead{color } & \colhead{0.7  } & \colhead{1.0} & \colhead{1.3}  }
\startdata
2.0  &  F9 V  &  +1.6  &   +0.8 &  +0.3 \\
2.5  &  F3 V  &  +1.1  &   +0.3 &  $-$0.2 \\
3.0  &  A5 V  &  +0.6  &  $-$0.2 & $-$0.7 \\

\enddata

\end{deluxetable}

\begin{figure}
\plotone{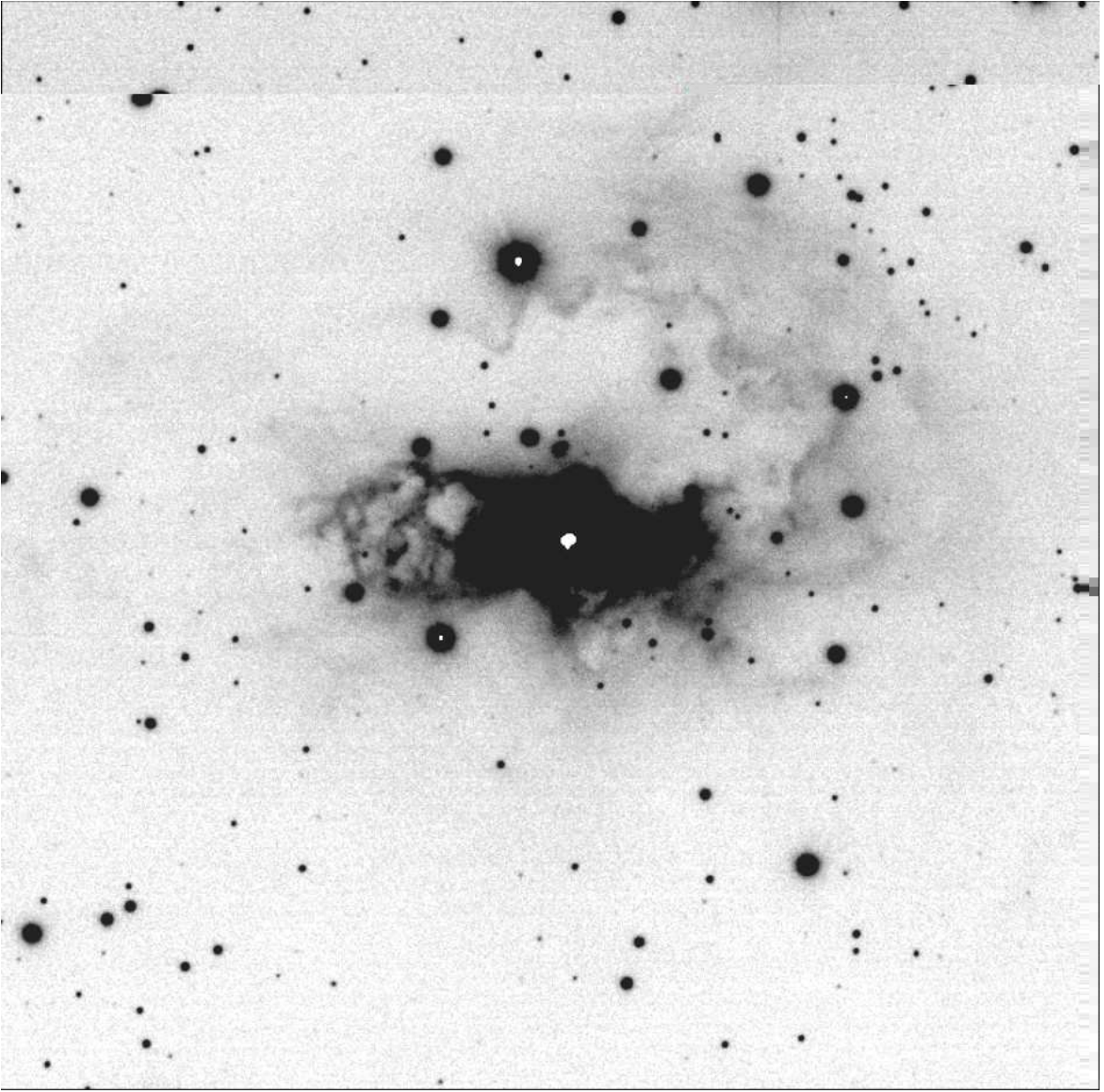}
\caption{ IC 2144 in a passband centered on H$\alpha$, obtained with the
Subaru telescope on 2006 Jan. 6, in 0$\farcs$6 to 0$\farcs$7 seeing.
The saturated spot indicates the location of MWC 778.  The dimensions are
about 204$\arcsec$ $\times$ 204$\arcsec$; north is at the top, east to the
left. }
\label{Fig. 1}
\end{figure}

\begin{figure}
\plotone{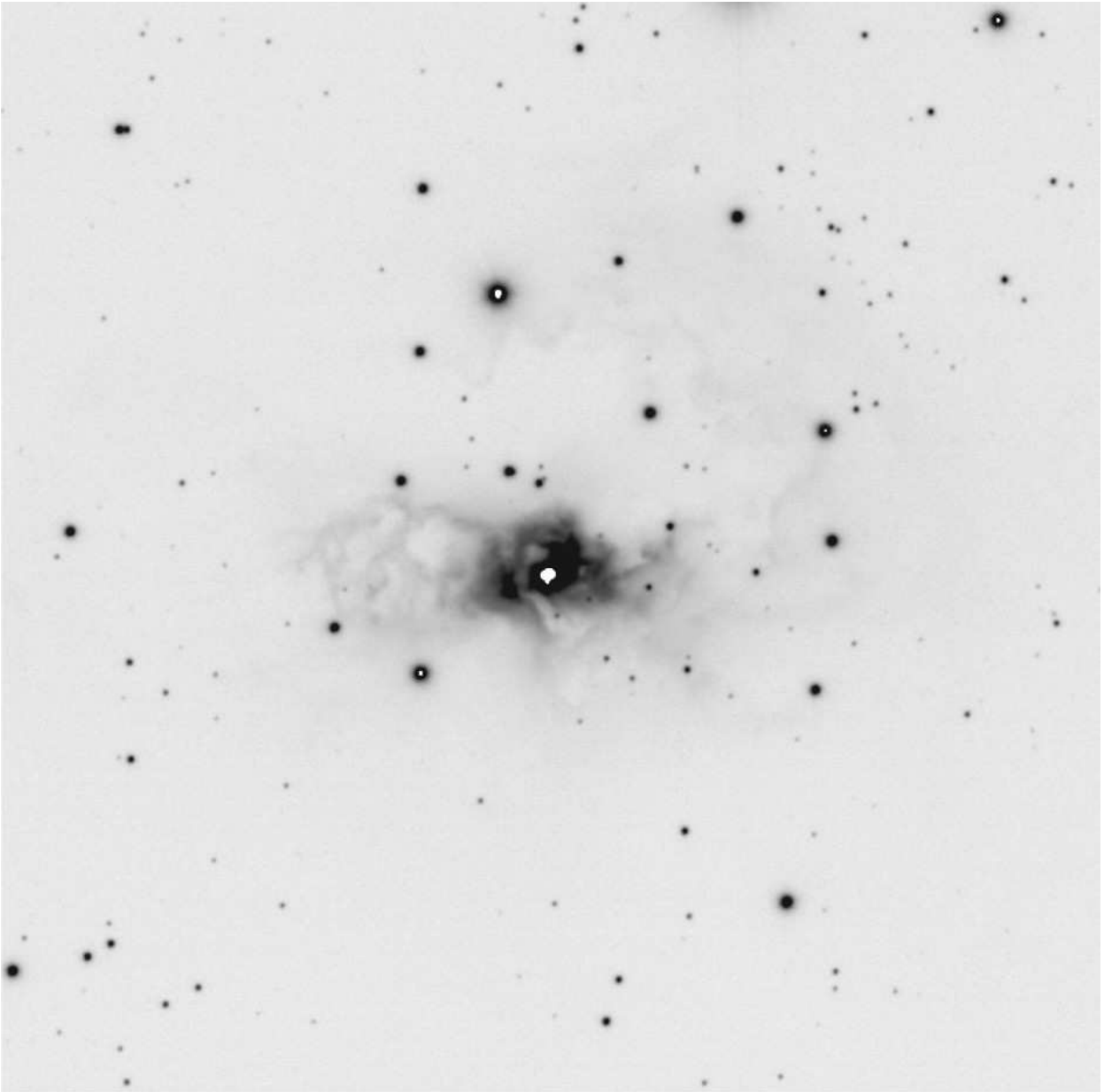}
\caption{ The same image as Fig. 1 but on a logarithmic intensity scale. }
\label{Fig. 2}
\end{figure}

\begin{figure}
\plotone{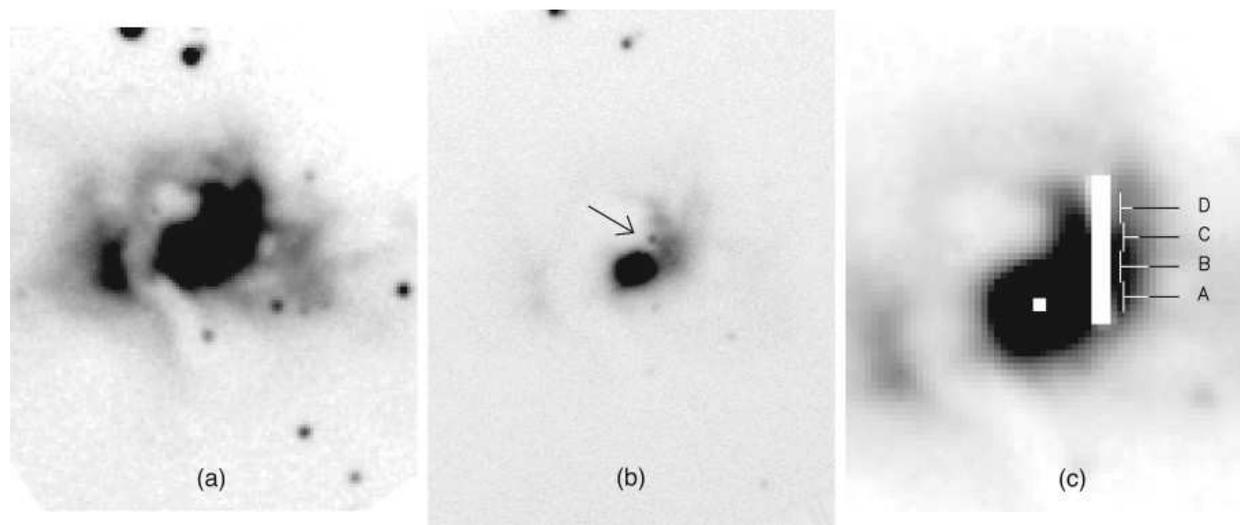}
\caption{ Three images of IC 2144 on the slit plate of the HIRES spectrograph;
Figs. 3a (left) and 3c (right) were obtained on 2003 December 13, in poor
seeing.  Fig. 3b (center)  was taken on 2004 November 21 in better
($\approx$1$''$) seeing. All have north at the top, east to the left. Fig 3a 
is printed so as to show the faint outer fringes of the nebulosity, while 3b
is on a logarithmic intensity scale, to better show the central star
(MWC 778) and the arcs of bright nebulosity extending to the northwest. The
arrow indicates the faint star (mentioned in the text) that is lost in the
bright nebulosity on the other exposures.  The field size is
35\arcsec $\times$ 42\arcsec in both cases.  Fig. 3c (right) is an
enlargement of the central region (field size 19$\arcsec$ $\times$ 22$\arcsec$)
showing the location of the slit when the spectrogram of the ``arc'' described
in \S\ 5 was obtained.  The letters indicate the locations of the slit
segments where individual spectra were extracted.  The small white square
marks the location of MWC 778.  }
\label{Fig. 3}
\end{figure} 

\epsscale{1.}
	
\begin{figure}
\plotone{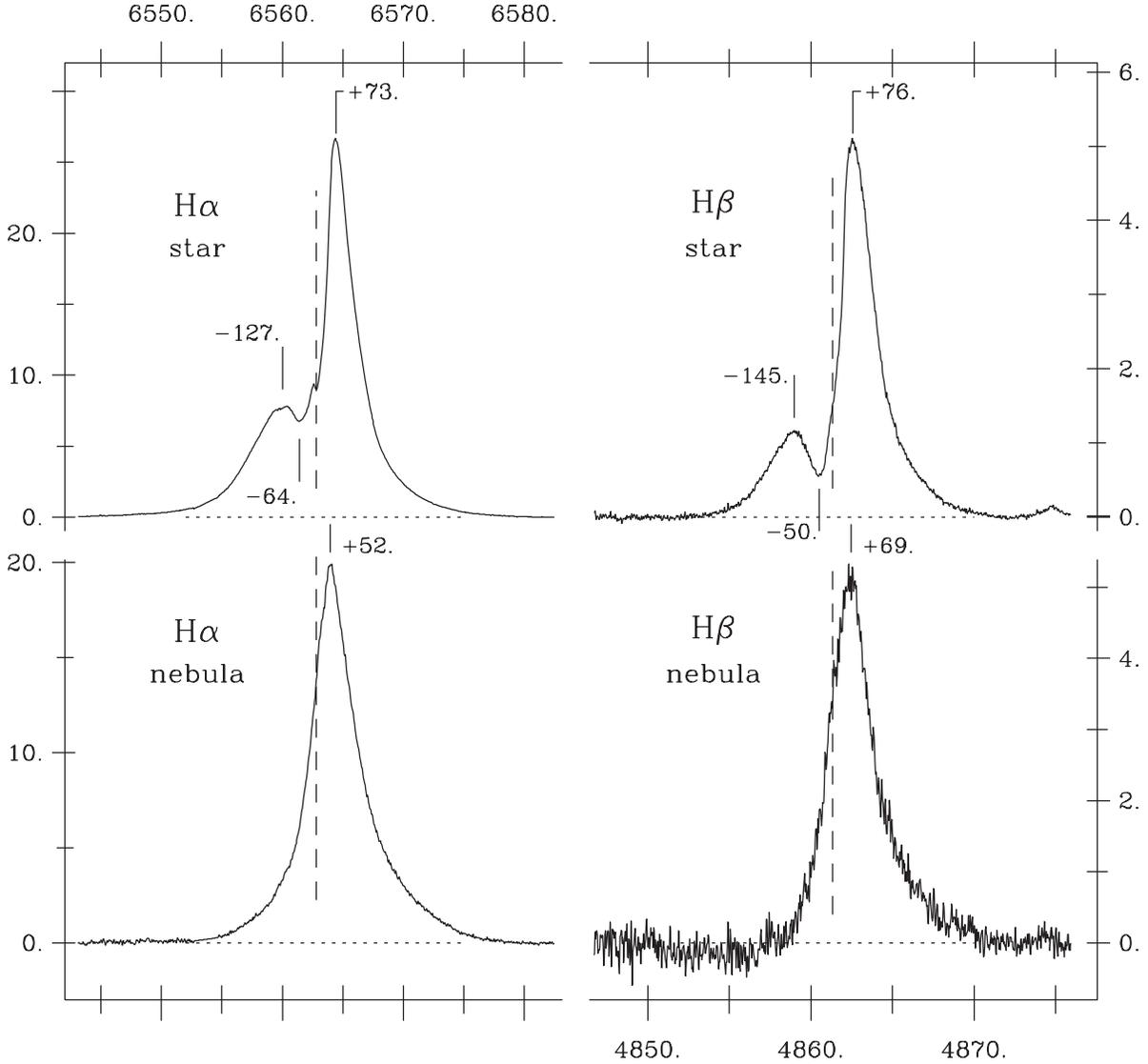}
\caption{Above:  the profiles of H$\alpha$ and H$\beta$ in MWC 778, from the
2003 HIRES observation.  Below: the same lines in the nebula;  the latter 
are the sums of the spectra extracted from the 4 individual slit segments.
The vertical dashed lines mark the zero velocity positions.
Both lines in the nebula are nearly symmetric and structureless,  while
the shortward absorption minimum in the star is absent in the nebula.  Also
the line peaks in the nebula are displaced shortward with respect to their
positions in the star.  These differences, as well as a less obvious mismatch
in the longward wing, are shown for H$\alpha$ in Fig. 10. The radial
velocities of several features (with respect to the rest velocity of the line)
are indicated.  The abscissae are in angstroms.  }
\label{Fig. 4}
\end{figure}

\begin{figure}
\includegraphics[height=7.0in,keepaspectratio=true,origin=c,angle=0.]
{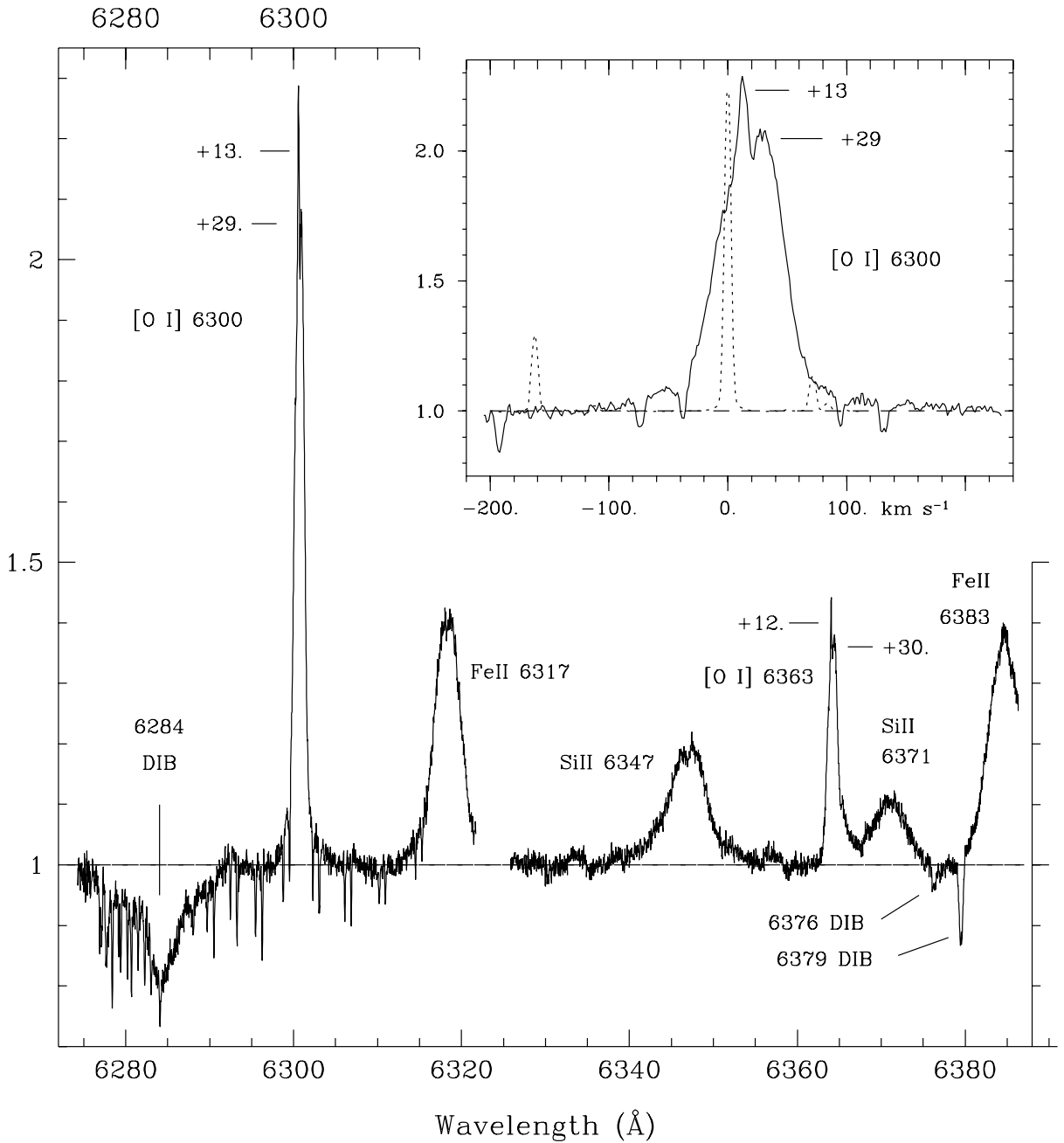}
\caption{ The 6280--6380 \AA\ region in MWC 778, from the exposure of 2004
November 21.  It illustrates the difference in widths between the emission lines
of $[$\ion{O}{1}] and the permitted lines of \ion{Si}{2} and \ion{Fe}{2}.  The
lines of $[$\ion{Ni}{2}$]$ and $[$\ion{Cr}{2}$]$ in the near infrared are
similiarly narrow.  The case of $[$\ion{Fe}{2}$]$ is discussed in the text. 
The $[$\ion{O}{1}] $\lambda$6300 line is expanded and plotted on a velocity
scale in the inset.  The dotted outline is of a thorium comparison line, to
show the instrumental resolution. }
\label{Fig. 5}
\end{figure}

\begin{figure}
\plotone{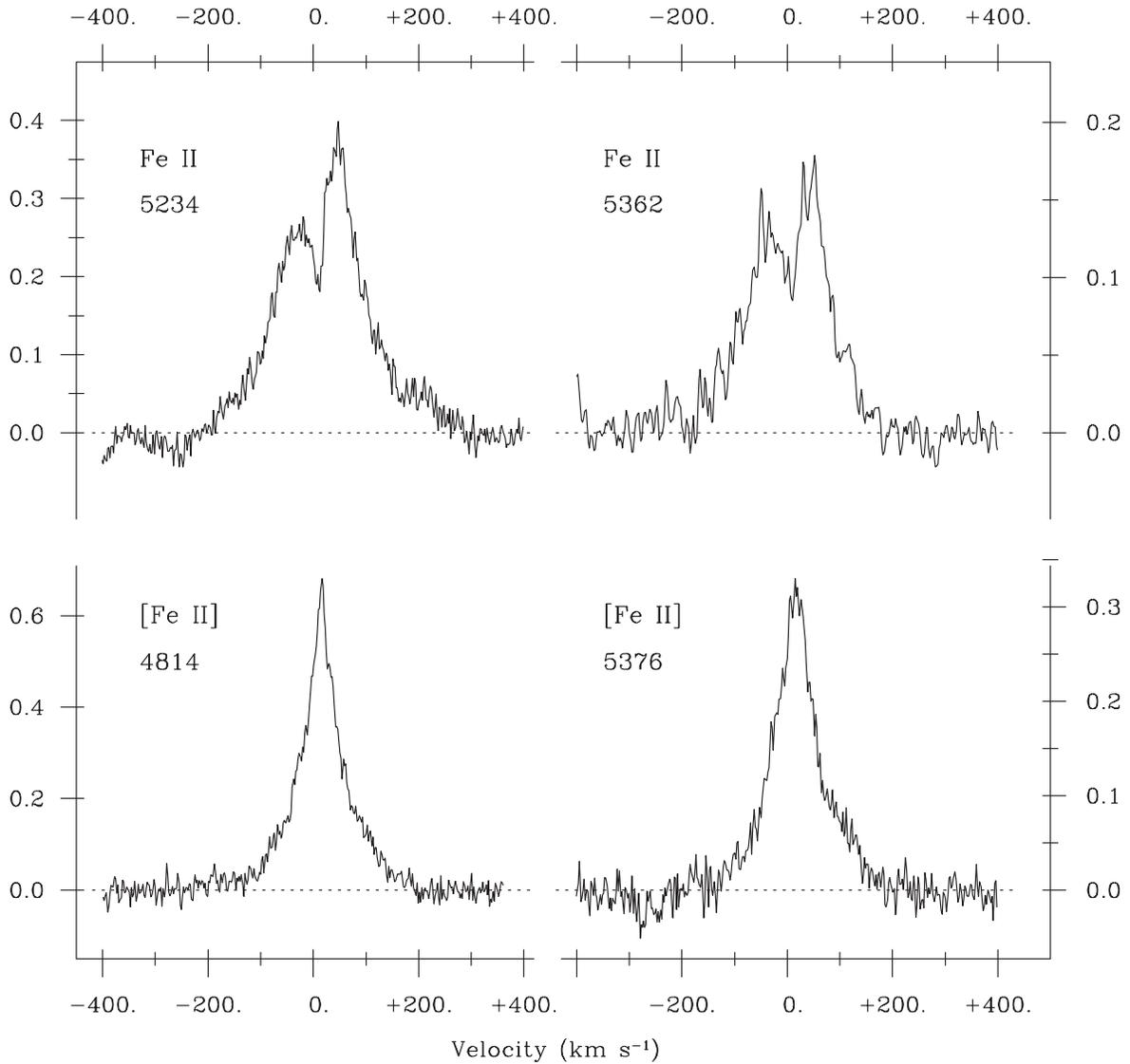}
\caption{  MWC 778: The profiles of two \ion{Fe}{2} emission lines (above) and
two [\ion{Fe}{2}] lines (below), showing that for \ion{Fe}{2}, the permitted
lines are always double and the forbidden lines always single---and lie at
an intermediate velocity. (The profile of $\lambda$5362 has been smoothed
slightly to reduce the noise.) }
\label{Fig. 6}
\end{figure} 

\begin{figure}
\plotone{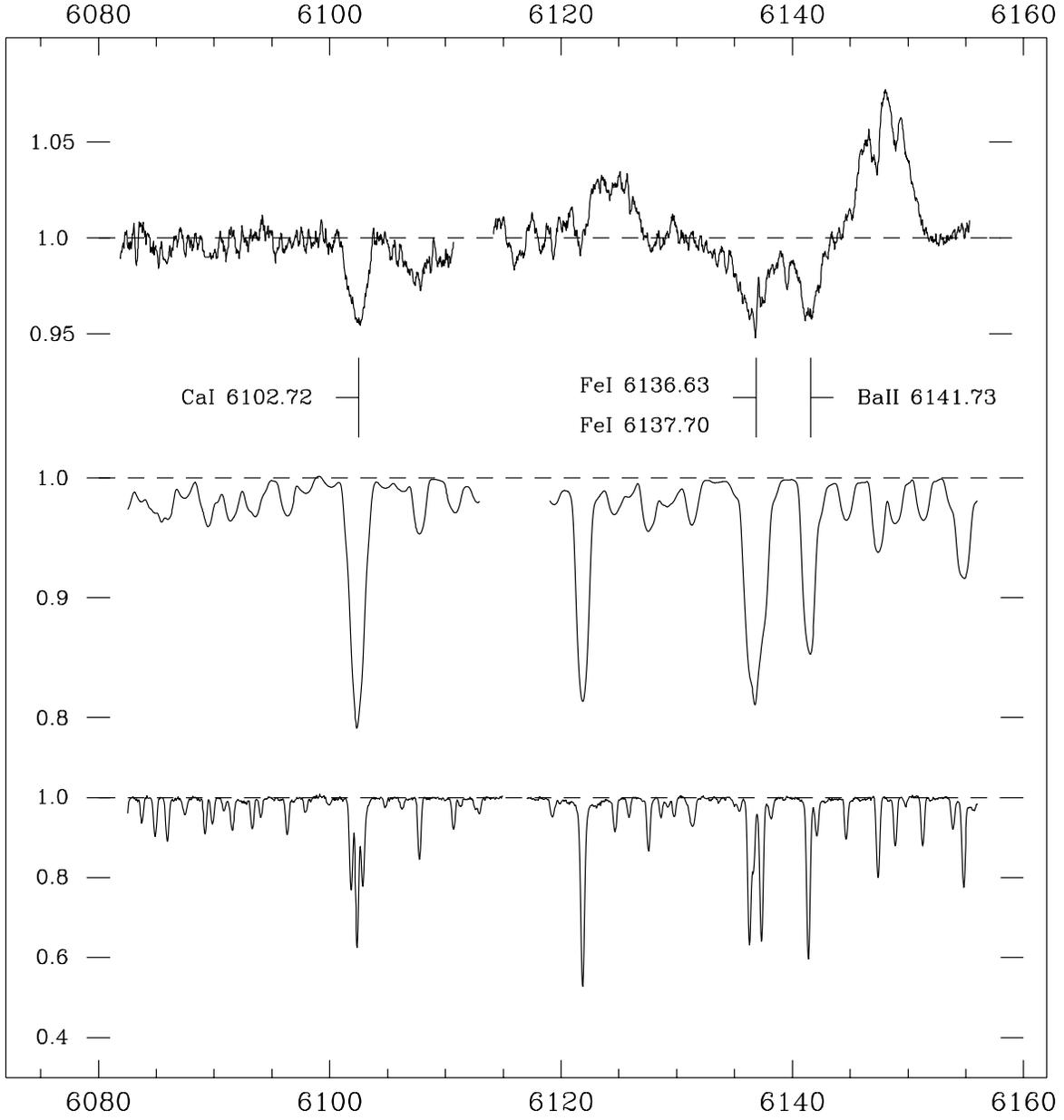}
\caption{ A comparison of the absorption spectrum of MWC 778 with that of
a normal G dwarf.  The 6080--6160 \AA\ region in MWC 778 (top), in
the G0\, V star HR 8314 (bottom), and in HR 8314 broadened
artificially to $v$ sin $i$ = 35 km s$^{-1}$ (middle). The main contributors
 to the absorption lines are identified.  Note that the vertical scales
of the three
spectra are very different;  they have been expanded vertically so
that the main features have comparable amplitudes.  The spectrum of MWC 778
has been shifted in wavelength to alignment with HR 8314.  The emission
lines in MWC 778 are \ion{Mn}{2} 6122.44 + 6125.85 \AA\ and
\ion{Fe}{2} 6147.73 + 6149.24 \AA.  The Mn\, II emission largely obliterates
the \ion{Ca}{1} 6122.22 \AA\ absorption line. }
\label{Fig. 7}
\end{figure}

\begin{figure}
\includegraphics[height=5.0in,keepaspectratio=true,origin=c,angle=0.]
{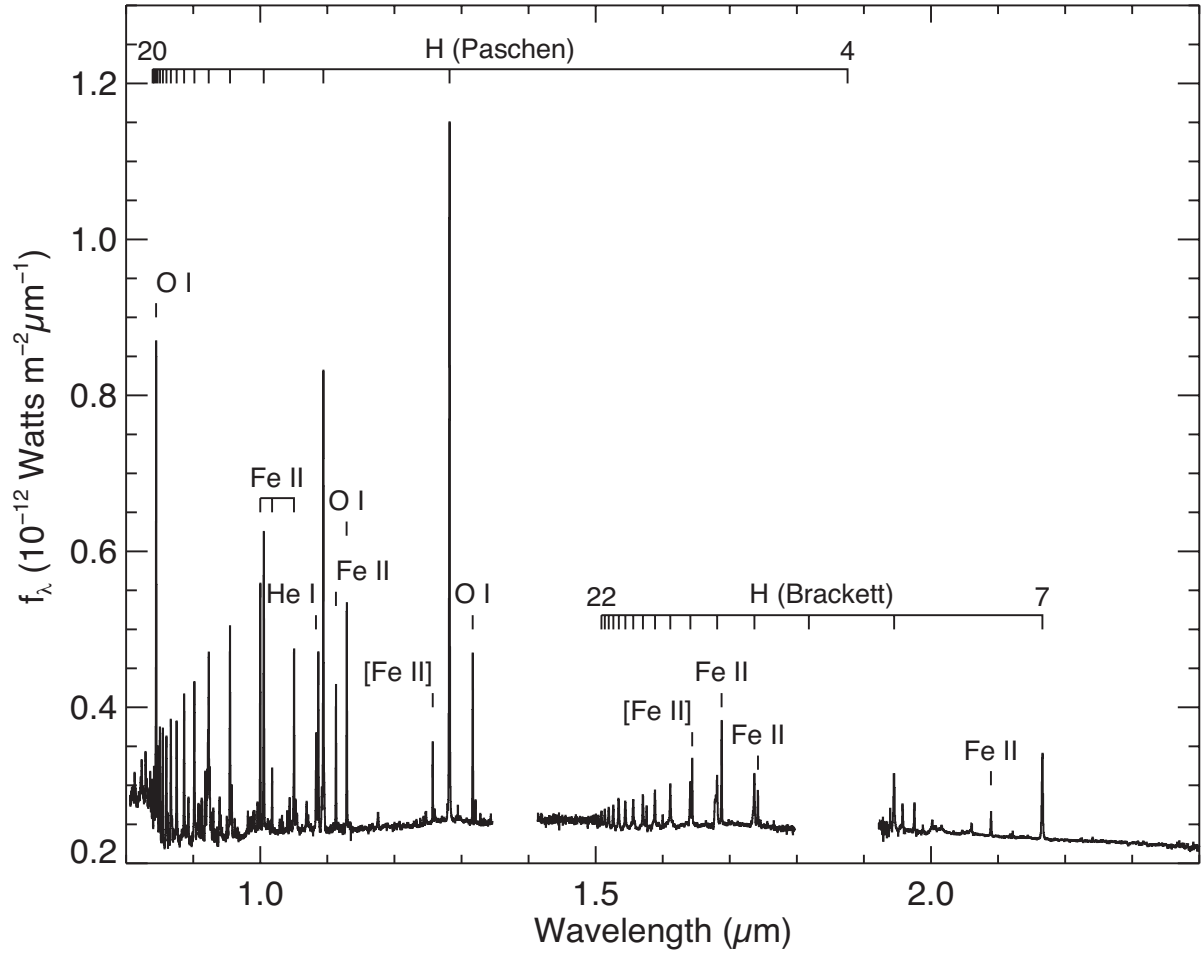}
\caption{Flux-calibrated NIR spectrum of MWC 778 obtained with SpeX on
the IRTF.  Regions of poor atmospheric transmission have been removed.
The strongest emission lines are identified.  The rise in flux seen at
0.82 $\mu$m is the Paschen jump. }
\label{Fig. 8}
\end{figure} 

\begin{figure}
\plotone{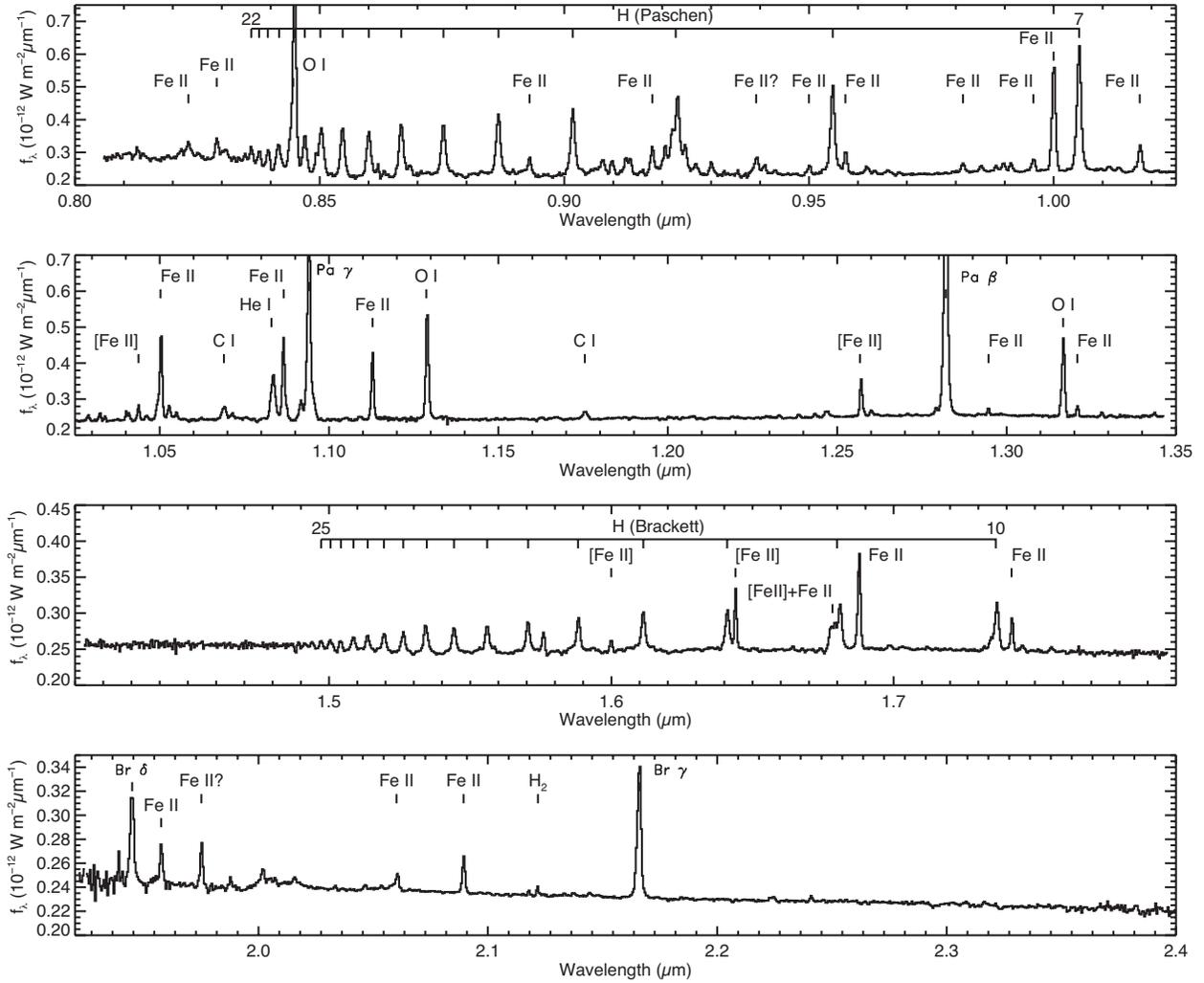}
\caption{The near-infrared spectrum of MWC 778, on an expanded scale,
showing the wealth of emission lines.  The strongest lines are identified.
The broad emission  seen at 2.00--2.02 $\mu$m is an artifact resulting
from the telluric correction procedure.}
\label{Fig. 9}
\end{figure}

\begin{figure}
\plotone{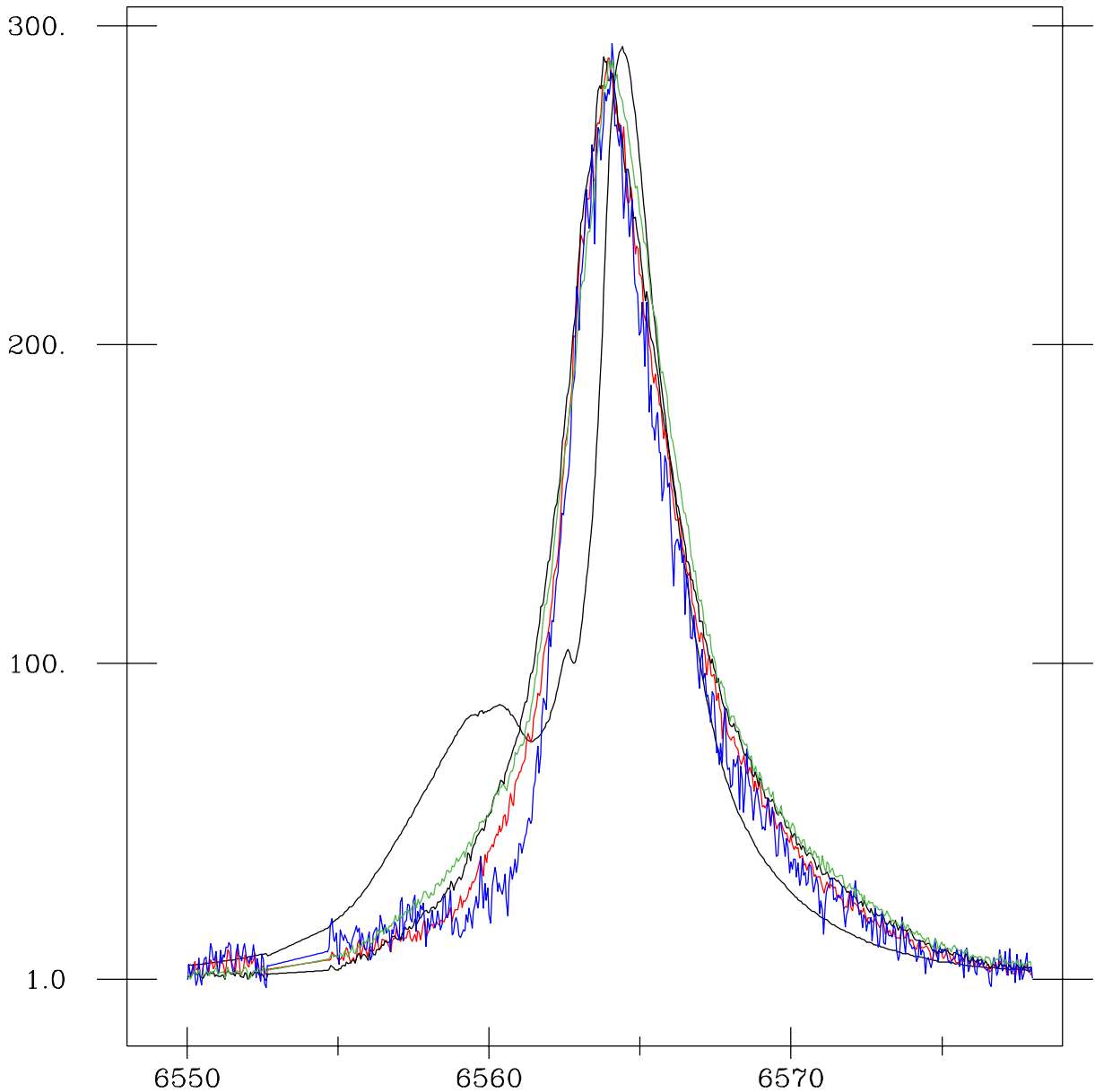}
\caption{  The profile of H$\alpha$ in MWC 778 (solid black line) superimposed
on the profile of H$\alpha$ extracted from four sections of the slit that
crossed the arc of nebulosity northwest of MWC 778 and that are identified
in Fig. 3c.  The colors indicate green = section A, black dotted = B, red = C,
blue = D.  The peak intensities have been set equal for this plot.  Note that
the shortward reversal and the asymmetry of the main peak of the star's
H$\alpha$ are not present in the spectra of the 'arc'.  Furthermore, the peak
in the star is about 20 km s$^{-1}$ longward of that in the nebula, and the
longward wings do not match.  There are also more subtle differences between
the H$\alpha$ profiles from the various slit sections. }
\label{Fig. 10}
\end{figure}

\newpage

\begin{figure}
\centering
\includegraphics[height=5.0in,keepaspectratio=true,origin=c,angle=0.]
{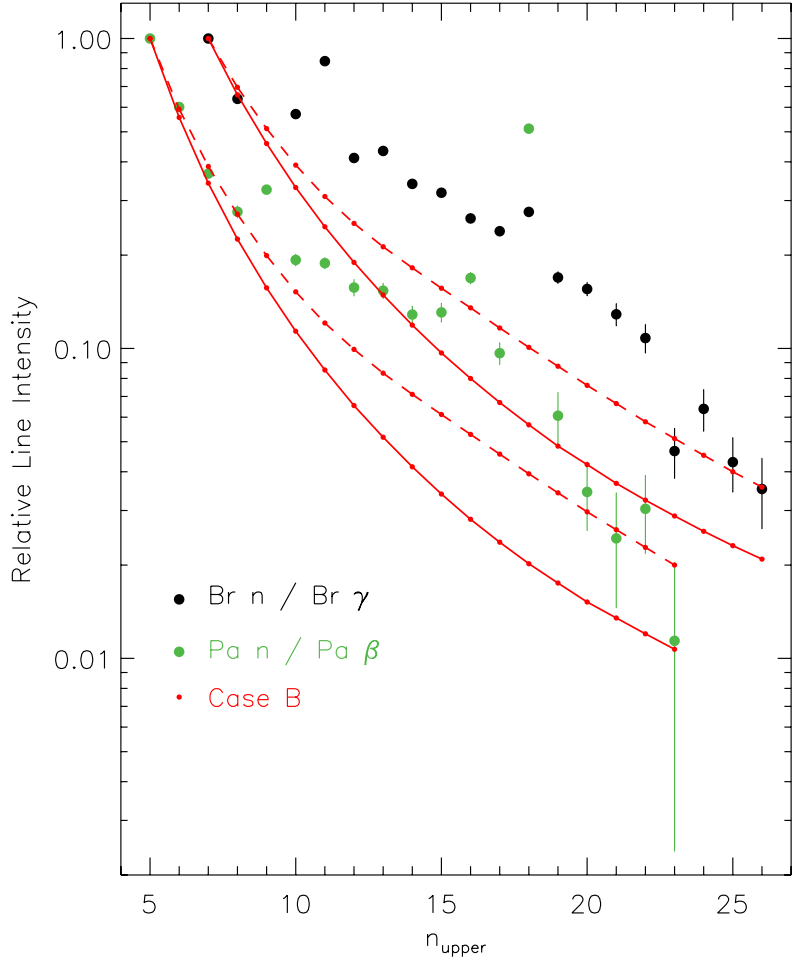}
\caption{Observed line intensities of the Paschen series, relative to the
strength  of Paschen $\beta$; and the Brackett series, relative to the
strength of Brackett $\gamma$, in the SpeX spectrum of MWC 778.  The solid
lines give the theoretical prediction for Case B from Hummer \& Storey
(1987) for $T_{e}$ =10$^{4}$ K and $n_{e}$ = 10$^{4}$ cm$^{-3}$
for the Br line ratio intensities (upper solid curve) and for the Pa lines 
(lower solid curve).  The dashed
lines represent the Case B predictions for $T_{e}$ = 10$^{4}$ K and $n_{e}$
 = 10$^{8}$ cm$^{-3}$ for the Br line ratio intensities (upper dashed
curve) and for the Pa lines  (lower dashed curve).  The observed
line ratios for large $n_{\rm upper}$ are
far stronger than predicted by Case B for any density value, which suggests
that the lines are optically thick.  Correction of the observed line
fluxes for extinction increases the deviations from the Case B prediction. }
\label{Fig. 11}
\end{figure} 
  
\begin{figure}
\plotone{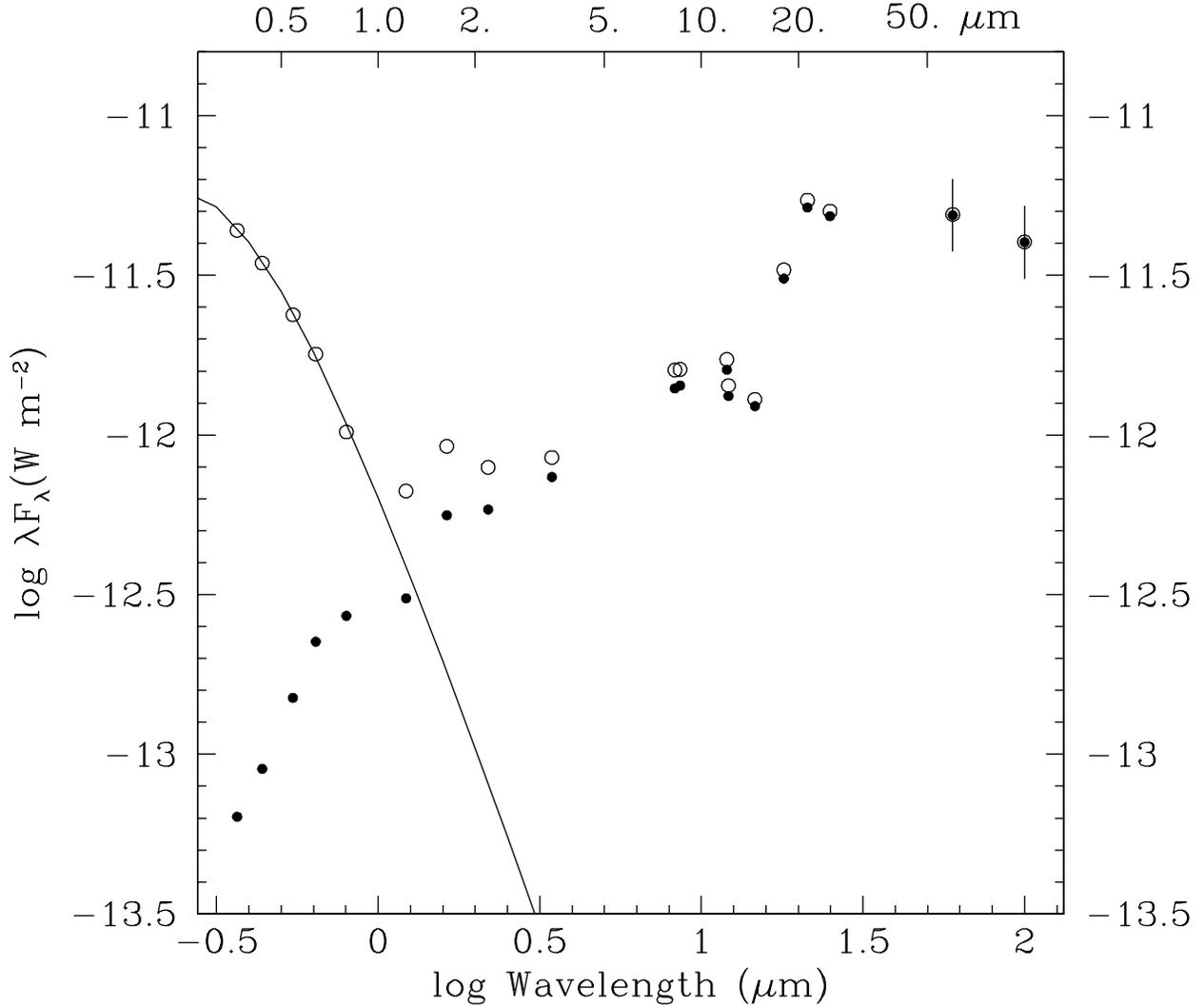}
\caption{The spectral energy distribution of MWC 778/IC 2144.  The solid points
are the observed values, constructed from the data of Table 1.  The open
circles are the same data corrected for normal reddening of $A_{V}$ = 3.0
 mag.  The solid line is the Planck distribution for T = 15,000 K,
shifted vertically to fit the optical points.
The vertical bars on the 60- and 100 $\mu$m points indicate 30\% errors. }
\label{Fig. 12}
\end{figure}

\begin{figure}
\includegraphics[height=7.0in,keepaspectratio=true,origin=c,angle=0.]
{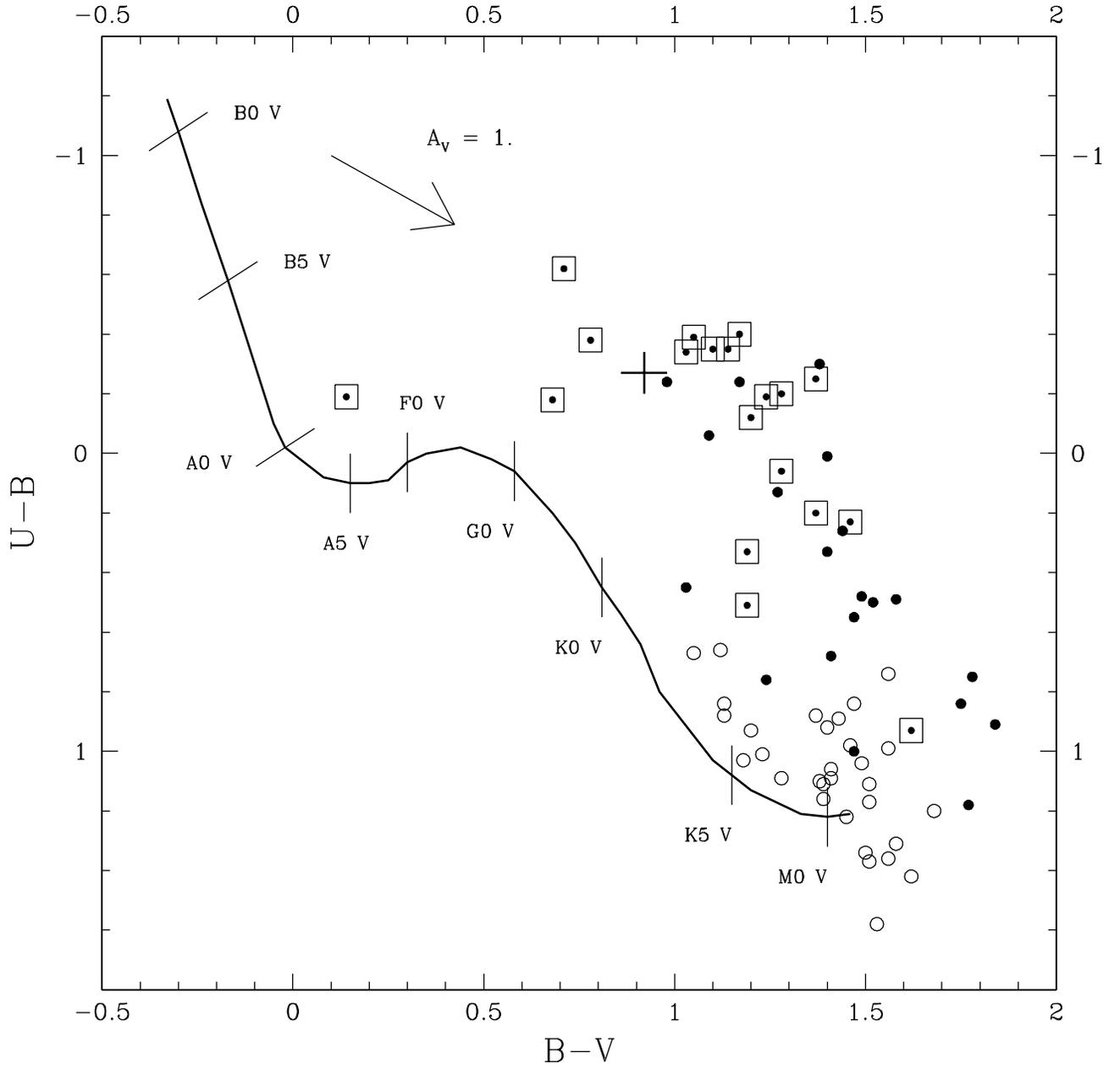}
\caption{The location in the $U-B$, $B-V$ plane of about 70 of the 
pre--main-sequence stars in the Tau-Aur clouds, from the compilation of
\citet{ken95}.  (Stars having errors $>$0.30 mag. in either coordinate are
not shown.)  The solid points in boxes are classical TTSs (CTTSs) having
$W(H\alpha)$ $\geq$ 60 \AA , solid points are CTTSs having $W(H\alpha$)
between 10 and 60 \AA , and open circles
represent WTTS ($W <$ 10 \AA ).  The W(H$\alpha$)'s are from \citet{herb88}.
The large cross marks the location of MWC 778. }
\label{Fig. 13}
\end{figure}

\end{document}